\documentclass[preprint]{aastex}
\usepackage{pdflscape}
\usepackage{multicol} 
\usepackage{epstopdf}
\usepackage{graphicx}

\shorttitle{Precise Radial Velocities of Low Mass Stars}
\shortauthors{GAO ET AL.}

\newcommand{\ms}{\hbox{m~s$^{-1}$}}

\begin{document}

\title{Retrieval of Precise Radial Velocities from Near-Infrared High Resolution Spectra of Low Mass Stars}

\author{Peter Gao\altaffilmark{1}, Plavchan P.\altaffilmark{2}, Gagn\'e J.\altaffilmark{3,4}, Furlan E.\altaffilmark{5}, Bottom M.\altaffilmark{6}, Anglada-Escud\'e G.\altaffilmark{7,8}, White R.\altaffilmark{9}, Davison C. L.\altaffilmark{9}, Beichman C.\altaffilmark{5}, Brinkworth C.\altaffilmark{5,10}, Johnson J.\altaffilmark{11}, Ciardi D.\altaffilmark{5}, Wallace K.\altaffilmark{12}, Mennesson B.\altaffilmark{12}, von Braun K.\altaffilmark{13}, Vasisht G.\altaffilmark{12}, Prato L.\altaffilmark{13}, Kane S. R.\altaffilmark{14}, Tanner A.\altaffilmark{15}, Crawford T. J.\altaffilmark{12}, Latham D.\altaffilmark{11}, Rougeot R.\altaffilmark{16}, Geneser C. S.\altaffilmark{2}, Catanzarite J.\altaffilmark{17}}

\altaffiltext{1}{Division of Geological and Planetary Sciences, California Institute of Technology, MC 150-21, 1200 East California Boulevard, Pasadena, CA 91125, USA; pgao@caltech.edu}
\altaffiltext{2}{Department of Physics, Missouri State University, 901 S National Ave, Springfield, MO 65897, USA}
\altaffiltext{3}{Department of Terrestrial Magnetism, Carnegie Institution of Washington, Washington, DC 20015, USA}
\altaffiltext{4}{Sagan Fellow}
\altaffiltext{5}{NASA Exoplanet Science Institute, California Institute of Technology, 770 S. Wilson Ave., Pasadena, CA 91125, USA}
\altaffiltext{6}{Division of Physics, Mathematics, and Astronomy, California Institute of Technology, Pasadena, CA 91125, USA}
\altaffiltext{7}{School of Physics and Astronomy, Queen Mary University of London, 327 Mile End Rd, E1 4NS, London, UK}
\altaffiltext{8}{Centre for Astrophysics Research, University of Hertfordshire, College Lane, AL10 9AB, Hatfield, UK}
\altaffiltext{9}{Department of Physics and Astronomy, Georgia State University, Atlanta, GA 30303, USA}
\altaffiltext{10}{National Center for Atmospheric Research, P.O. Box 3000, Boulder, CO 80307}
\altaffiltext{11}{Institute for Theory and Computation, Harvard-Smithsonian Center for Astrophysics, 60 Garden Street, Cambridge, Massachusetts 02138 USA}
\altaffiltext{12}{Jet Propulsion Laboratory, California Institute of Technology, 4800 Oak Grove Drive, Pasadena, CA 91125, USA}
\altaffiltext{13}{Lowell Observatory, West Mars Hill Road, Flagstaff, AZ 86001, USA}
\altaffiltext{14}{Department of Physics \& Astronomy, San Francisco State University, 1600 Holloway Avenue, San Francisco, CA 94132, USA }
\altaffiltext{15}{Mississippi State University, Department of Physics \& Astronomy, Hilbun Hall, Starkville, MS 39762, USA}
\altaffiltext{16}{(ESA, European Space Research and Technology Centre)}
\altaffiltext{17}{NASA Ames Research Center, MS 245-3, P.O. Box 1, Moffett Field, CA 94035-0001, USA}

\begin{abstract}

Given that low-mass stars have intrinsically low luminosities at optical wavelengths and a propensity for stellar activity, it is advantageous for radial velocity (RV) surveys of these objects to use near-infrared (NIR) wavelengths. In this work we describe and test a novel RV extraction pipeline dedicated to retrieving RVs from low mass stars using NIR spectra taken by the CSHELL spectrograph at the NASA Infrared Telescope Facility, where a methane isotopologue gas cell is used for wavelength calibration. The pipeline minimizes the residuals between the observations and a spectral model composed of templates for the target star, the gas cell, and atmospheric telluric absorption; models of the line spread function, continuum curvature, and sinusoidal fringing; and a parameterization of the wavelength solution. The stellar template is derived iteratively from the science observations themselves without a need for separate observations dedicated to retrieving it. Despite limitations from CSHELL's narrow wavelength range and instrumental systematics, we are able to (1) obtain an RV precision of 35 $\ms$ for the RV standard star GJ 15 A over a time baseline of 817 days, reaching the photon noise limit for our attained SNR, (2) achieve $\sim$ 3 $\ms$ RV precision for the M giant SV Peg over a baseline of several days and confirm its long-term RV trend due to stellar pulsations, as well as obtain nightly noise floors of $\sim$2\textendash 6 $\ms$, and (3) show that our data are consistent with the known masses, periods, and orbital eccentricities of the two most massive planets orbiting GJ 876. Future applications of our pipeline to RV surveys using the next generation of NIR spectrographs, such as iSHELL, will enable the potential detection of Super-Earths and Mini-Neptunes in the habitable zones of M dwarfs. 
\end{abstract}

\section{Introduction}\label{sec:intro}

The radial velocity (RV) technique has been extremely successful in the detection and confirmation of exoplanets \citep[]{latham1989,mayor1995}. In keeping with our pursuit of Earth-like worlds, the majority of RV targets have been FGK stars similar to our Sun \citep[]{akeson2013}. However, this neglects M dwarfs, which make up 75$\%$ of the stars in the Solar neighborhood \citep[]{henry2006}. M dwarfs' habitable zones (HZs) are also at smaller semi-major axes due to their lower luminosities; coupled with their lower masses, this results in larger radial velocity perturbations from a companion planet in the HZ than for FGK stars. Surveys of close-by early- and mid-M dwarfs have reached RV precisions down to $\sim$ 2 $\ms$, sufficient to detect Super-Earths in these stars' HZs \citep[]{zechmeister2009,bonfils2013}; by comparison, a twenty-fold increase in RV precision is needed to find Earth-mass planets in the HZs of Sun-like stars.

The detection of planets around M dwarfs is hampered by two important factors, however. First, K and M stars are generally more active than G stars, with late-M dwarfs being the most active \citep[]{west2004,basri2010}. Stellar activity, such as starspots, can produce false positive planet signatures by introducing additional power at the frequency of the rotational period of the star and aliases thereof in the Doppler measurements \citep[]{robertson2015,vanderburg2015}. Second, the low luminosity of M dwarfs at optical wavelengths makes it difficult to obtain high signal to noise (SNR) measurements. A logical solution to these problems is to observe at near infrared (NIR) wavelengths. Not only do M dwarfs emit most of their light in the NIR, the temperature contrast between starspots and the rest of the chromosphere is also reduced, decreasing the periodic modulation of chromospheric activity on the Doppler measurements \citep[]{martin2006,reiners2010,mahmud2011,crockett2012,anglada2013}. This effect can help lower the rate of false positive optical planet detections by discriminating actual planets from the effects of spots \citep[]{huelamo2008,prato2008}. This also allows for NIR RV surveys to focus on younger stars, which tend to be more active. 

Several NIR RV surveys of M dwarfs have been conducted recently. \citet[]{rodler2012} used the NIRSPEC spectrograph at the Keck II Telescope to observe eight late-M dwarfs in the $J$ band ($\sim$ 1.1\textendash 1.4 $\mu m$) and obtained an RV precision of 180\textendash 300 $\ms$. Meanwhile, \citet[]{bailey2012} and \citet[]{blake2010} used NIRSPEC in the $K$ band ($\sim$ 2.0\textendash 2.4 $\mu m$) to obtain an RV precision of 50 $\ms$ for mid- and late-M dwarfs. \citet[]{tanner2012} also used NIRSPEC in the $K$ band to observe a sample of late-M dwarfs, with a resulting RV precision of 45 $\ms$. By comparison, simulations have shown that an RV precision of 25\textendash 30 $\ms$ for M dwarfs with $T_{eff} <$ 2200 K are possible in the NIR with spectrograph resolutions R $\sim$ 25000, similar to that of NIRSPEC in the $J$ and $K$ bands \citep[]{mclean1998}, though instrumental effects and systematic errors are not taken into account in this estimate \citep[]{rodler2011}. For earlier M dwarfs ($T_{eff} \sim$ 3500 K) and a higher resolution spectrograph (R $>$ 60000), $\sim$ 10 $\ms$ may be possible in the H and K bands \citep[]{reiners2010,bottom2013}. Even so, these results are more than an order of magnitude larger than the $\sim$ 1 $\ms$ RV precision limit of current optical RV surveys \citep[e.g.,][]{dumusque2012}. 

The lower RV precision of NIR RV surveys is partly attributable to the lack of a wavelength calibration method on par with those of optical surveys. Precise wavelength calibration is currently obtained either through extreme environmental stabilization (e.g. HARPS), or the use of a simultaneous common optical path wavelength reference. The latter process involves the comparison of the observations with a set of ``standard'' well-calibrated spectra. In the previously mentioned NIR RV studies, telluric lines were used for calibration, which were imprinted on the incoming star light from its passage through the Earth's atmosphere; comparisons between the observed telluric lines with models or observations of telluric spectra then provide a wavelength solution. In optical surveys, precise wavelength calibration is realized by placing an Iodine gas cell in the telescope optical path \citep[]{marcy1992,butler1996}. As the incoming star light passes through the gas cell, it picks up the spectral signatures of its gas. Comparing these observed spectra with a precisely measured absorption spectrum of the gas cell at higher spectral resolution than the astronomical spectrograph (R $\sim$ 500,000 obtained with a Fourier Transform Spectrometer) then yields a wavelength solution. Placing the gas cell in the optical path of the telescope also ensures that the spectral lines are obtained simultaneously and under the same physical conditions (i.e. temperature, pressure, humidity) as the instrument and the telescope, which is not the case for telluric lines. Similar gas cells have recently been developed for the NIR, such as the ammonia gas cell used by \citet[]{bean2010} on the CRIRES spectrograph at the VLT to obtain an RV precision of $\sim$ 5 $\ms$ for late-M dwarfs. Gas cells filled with hydrocarbons, chlorocarbons, HCl, and isotopologues of HCN, C$_2$H$_2$, CO, and CH$_4$ have also been considered \citep[]{mahadevan2009,valdivielso2010,anglada2012,plavchan2013}. 

An additional issue with current NIR RV surveys is that they rely on large (8-10 m aperture) telescopes, which generally have fewer nights available for RV monitoring. In contrast, smaller telescopes can provide the high cadence necessary for precise RV observations. For example, \citet[]{prato2008,crockett2011,crockett2012,davison2015} were able to use the smaller (3 m aperture) NASA Infrared Telescope Facility (IRTF), in conjunction with telluric line and emission lamp wavelength calibration, to obtain RV precision of 60\textendash 130 $\ms$ for M dwarfs in the K band using the CSHELL spectrograph, which has a spectral resolution R $\sim$ 46000 \citep[]{greene1993}. In addition, many of the instruments currently in development for dedicated NIR RV surveys will be installed on smaller aperture telescopes, e.g. CARMENES/Calar Alto Astronomical Observatory \citep[]{quirrenbach2010}, SPIRou/Canada France Hawaii Telescope \citep[]{santerne2013}, and iSHELL/NASA Infrared Telescope Facility \citep[]{rayner2012}. The two former instruments will rely on extreme stabilization, whereas the latter will use gas cells for stable wavelength calibration. It is essential that we prepare for these next generation NIR RV surveys by focusing on data reduction and analysis methods that can process their results. 

In this paper, we describe in detail the NIR RV extraction pipeline used in our RV survey of M dwarfs conducted using the NASA IRTF CSHELL spectrograph \citep[]{greene1993}, and presented in \citet[]{anglada2012,plavchan2013,gagne2015}. This pipeline uses the ``Grand Solution'' approach, which derives the stellar template, wavelength solution, and RVs simultaneously \citep[]{valenti2010} in an iterative fashion. This differs from other methods used in some previous RV surveys, such as substituting imperfect stellar spectral models as stellar templates. Thus, our pipeline offers a novel way to extract RVs from raw spectra from past, present, and future observation campaigns in the NIR. We test our pipeline on the RV standard GJ 15 A to assess the RV precision obtainable for early-M dwarfs, as well as the M Giant SV Peg to evaluate the pipeline's precision at the high SNR limit. We also test it on the planet host GJ 876 to validate its planet detection capabilities. 

We outline our observations of GJ 15 A, SV Peg, and GJ 876 in ${\S}$\ref{sec:observation} and describe our data reduction process in ${\S}$\ref{sec:reduction}. In ${\S}$\ref{sec:pipeline} we detail our RV extraction pipeline and in ${\S}$\ref{sec:results} we present our results. We summarize our work and discuss future prospects in ${\S}$\ref{sec:discussion}.

\section {Observations}\label{sec:observation}

Three targets were picked from our NIR RV survey to test the RV stability, high SNR behavior, and planet detection capabilities, of the RV pipeline presented here. These targets are summarized in Table \ref{table:starproperties}. GJ 15 A was chosen as a RV standard star due to its brightness, as well as its established RV scatter of a few $\ms$ due mostly to an orbiting exoplanet \citep[]{endl2006,howard2014}, which is below our estimated level of precision \citep[]{anglada2012}. SV Peg is an M giant star that has already been used as a reliable high SNR target in \citet[]{anglada2012}, and therefore will be used for our high SNR tests as well. GJ 876 has four confirmed planets, with the most massive two causing RV amplitudes $\sim$ 200 $\ms$, which should be easily detectable with our pipeline \citep[]{delfosse1998,marcy1998,marcy2001,rivera2005,rivera2010}. 


Data were taken on the 3 m telescope at NASA IRTF from September 2010 to August 2012 for GJ 15 A and GJ 876, and to August 2011 for SV Peg\footnote{No public archive currently exists for NASA IRTF data, but spectra used in this study can be made available upon request to the authors.}. Table \ref{table:observations} gives the minimum, maximum, median, and total SNR per pixel of the observations of each night, as well as the number of observations ($N_{obs}$) obtained that night for the three targets. Single order spectra in the $K$ band (2.309\textendash 2.315 $\mu m$) were obtained using the CSHELL spectrograph with a 0\farcs5 slit and a spectral resolution R $\sim$ 46000. A circular variable filter (CVF) was used for order selection. A methane isotopologue ($^{13}$CH$_{4}$) gas cell with 90\% continuum throughput was placed into the beam to achieve wavelength calibration \citep[]{anglada2012,plavchan2013}. To set the wavelength scale at the start of observations, we observed A-type stars with the gas cell in the optical path of the telescope and matched the deepest gas cell line to pixel column 179 on the CSHELL detector. For the brightest A stars a latent image of the target was sometimes left on the detector after switching from imaging mode (used to position the target in the slit) to spectroscopic mode; these spectra were flagged during the data reduction process and discarded. Around 15 flat fields and darks were taken in spectroscopic mode at the end of each night of observations, each with an exposure time of 20 s. Nodding along the slit was used during the September 2010 run, but only the ``A'' nod position was used here, as systematic errors were introduced into the RVs upon usage of different trace positions. Subsequent runs did not use nodding, as it was deemed unnecessary and detrimental to RV precision. Typical integration times were tuned to avoid nonlinear detector regimes ($>$2000 ADU per exposure) and to account for seeing variations, ranging from 60\textendash 200 s and 180\textendash 240 s for GJ 15 A and GJ 876, respectively, allowing for SNR per pixel $\sim$ 30 for each raw spectrum and $\sim$ 200 after $\sim$ 2 hours of combined integration. Seeing usually varied between 0\farcs5 and 1\farcs5. Exposure times of 5 s were sufficient to obtain SNR per pixel $\sim$ 100 for each spectrum of SV Peg under most seeing conditions. By comparison, the limiting magnitude of CSHELL in the K band ($\sim$ 2.2 $\mu$m) for SNR per resolution element of 10, resolving power of 21500 (1\farcs  slit), a total observing time of 1 hour, and a single exposure integration time of 120 s is 12.8 \citep[]{greene1993}. Converting from this standard to our case is difficult without a PSF model, which varies considerably for our observations. 

The time baseline of the observations analyzed in this work for GJ 15 A and GJ 876 are 817 and 707 days, respectively. For SV Peg, which is variable on a time scale $\sim$ 145 days with RV amplitude $\sim$ 1.5 km s$^{-1}$\citep[]{hinkle1997,lebzelter2002,winnberg2008}, we focus on intra-night and intra-run RV stability only. 

\section{Data Reduction}\label{sec:reduction}

All spectra were extracted consistently with a custom \emph{interactive data language} (IDL) pipeline. We found that our ability to achieve RV measurements at precisions $<$ 55 $\ms$ strongly relies on (1) a careful construction of the flat fields, (2) efficient detection and correction of bad pixels to prevent them from contaminating the extracted spectra, and (3) a correction of the instrumental fringing in the individual flat fields, which occurred at a $\sim $ 0.2\textendash 0.6\% level and are caused by the CSHELL CVF. Instrumental fringing also affects individual science images; however, it is not possible to efficiently correct them directly due to low illumination of most of the detector. For this reason, residual fringing was taken into account in our spectral model when radial velocities are extracted from the reduced spectra (see ${\S}$\ref{sec:pipeline}).

\begin{figure}[p]
\centering
\includegraphics[width=0.8 \textwidth, clip=true,trim=0cm 0cm 0cm 0cm]{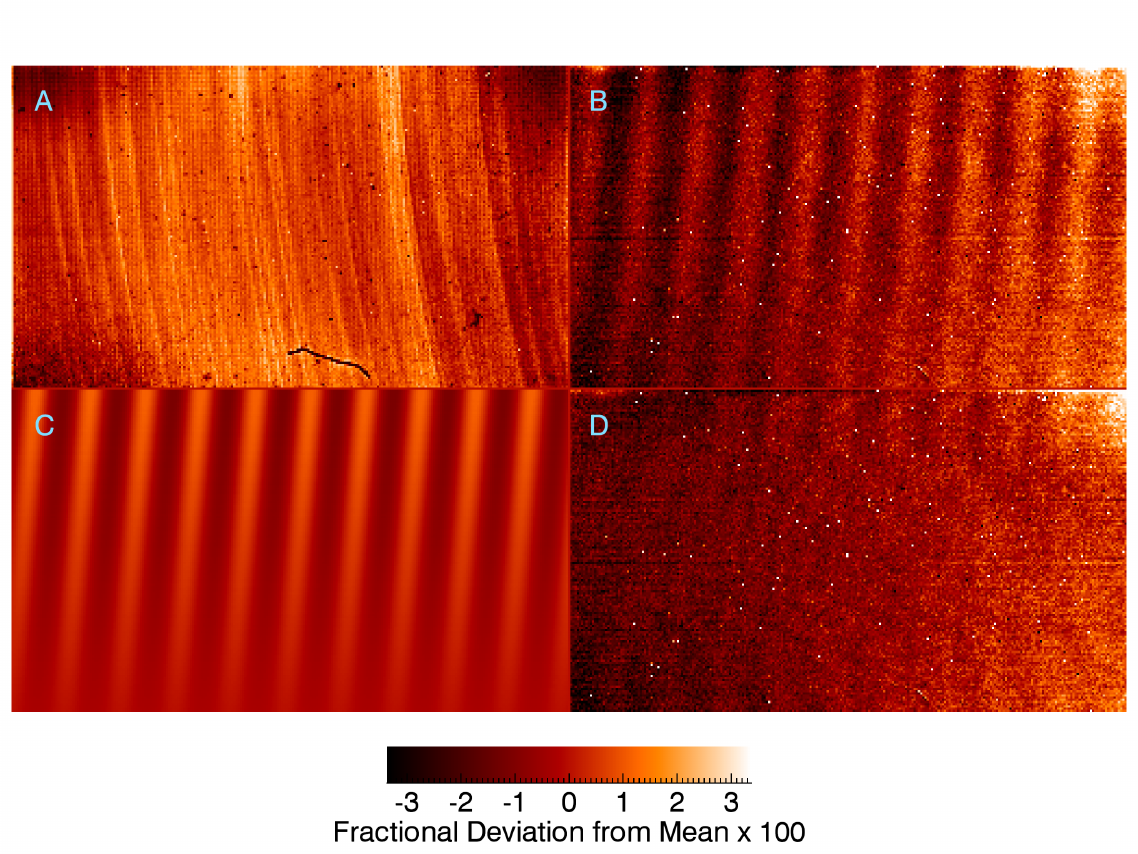}  
\caption{Removal of sinusoidal fringing from spectroscopic flat fields. (A) The master flat field generated from averaging all flat fields obtained in the 2010\textendash 2012 survey. (B) Result of dividing a median-combined nightly flat field by the master flat field. (C) The 2D fringing model fit to B. (D) Same as B, but after dividing out C to correct for the fringing.}
\label{fig:flatfrng}
\end{figure}

Combined spectroscopic flat field images were created for every night by median-combining typically 15 individual flats with exposure times of 20 seconds, from which a dark frame was subtracted. Removal of the instrumental fringing pattern in the flat fields is shown in Figure \ref{fig:flatfrng}. Fringing subtraction was done by first creating a fringing-free flat field (panel A) by average-combining a large number of per-night flat fields; fringing gets averaged out since its phase and amplitude vary randomly across observing nights, leaving behind only the permanent detector response. This fringing-free flat field was subsequently divided from each combined flat field (one per night) to make 2D fringing patterns apparent (panel B). We median-combined individual columns of this image to obtain a 1D fringing pattern. A Levenberg-Marquardt least-squares algorithm was then used to fit the 1D fringing pattern with a shifted and renormalized interference function of the form

\begin{equation}\label{eq:fringemodel}
	\mathcal{I}_1(P_o) = \frac{A_f}{n}\left( \frac{1}{1+\mathcal{F}\sin^2\left(\omega_p P_o +\phi \right)} - 1\right) + A_f + \mathcal{I}_0 + s P_o, \\
	\mbox{where}\ n = \frac{1}{2}\left(1-\frac{1}{1-\mathcal{F}}\right)
\end{equation}

\noindent is a normalization factor, $A_f$ is the amplitude of the interference pattern, $\mathcal{F}$ is the \emph{finesse} parameter, $P_o$ is the pixel grid on the detector, $\omega_p$ is the spatial frequency (in units of pixel$^{-1}$), $\phi$ is the phase, and $\mathcal{I}_0$ and $s$ allow for a linear slope in the flux. The resulting parameters are then used as starting estimates to fit the full 2D fringing pattern using the following model :

\begin{equation}\label{eq:fringemodel2}
	\mathcal{I}_2(P_o,y) = \left[\frac{1}{n}\left( \frac{1}{1+\mathcal{F}\sin^2\left(\omega_p P_o +\phi+\phi_y  y \right)} - 1\right) + 1\right]  \left(A_f+A_{f,y} y\right) + \mathcal{I}_0 + s_p P_o + s_y y + s_{p,y} P_o y
\end{equation}

\noindent where $y$ is the column number on the detector, perpendicular to $P_o$, $\phi_y$ allows for a linear phase shift with respect to rows (resulting in tilted interference fringes), $A_{f,y}$ allows for the amplitude to vary with rows and $\mathcal{I}_0$, $s_p$, $s_y$ and $s_{p,y}$ allow for a 2D plane to be fit to the residual flux. The resulting 2D fringing pattern (panel C), from which the linear plane fit was subtracted, $\mathcal{I}_2(P_o,y)+1-\left(s_p P_o + s_y y + s_{p,y} P_o y \right)$ was finally divided from the combined flat field to obtain a corrected flat field to be used in the data reduction process (panel D, multiplied by the fringing-free flat field). Though the fringing is not fully removed at the top and bottom edges of the image, the target spectral trace is always located at the center of the image, and so our technique is sufficient.

Individual science images were read separately and divided by the corrected flat field of the corresponding observing night. We cross-correlated a median spatial profile of the spectrum at each spectral position to determine a linear solution to the trace tilt and corrected it by interpolating in the spatial direction only. This step was repeated three times iteratively.

We re-constructed a median spatial profile of the straightened trace and fitted a Moffat profile, defined by $A_m/((P_o-P_{o}')^2/\sigma_m^2+1)^{\gamma}$ where $A_m$ is the amplitude, $P_{o}'$ the central pixel position, $\sigma_m$ the characteristic width, and $\gamma$ an index that controls the size of the wings \citep[]{moffat1969}. We used the resulting fitting parameters to set all pixels to zero at positions $|P_o-P_{o}'| > \sigma_m / \sqrt{2\gamma+2.3}$ in the median spatial profile. The median spatial profile was then used in combination with a standard optimal extraction procedure \citep[]{horne1986,massey2013} to extract the spectrum a first time.

The resulting spectrum and spatial profile were then used to reconstruct a smooth synthetic 2D spectrum. We subtracted this synthetic spectrum from the raw science image, and created a map of small scale structures in the spatial direction by taking the minimum value between the absolute values of the upwards and downwards vertical (i.e., spatial direction) derivatives in the resulting difference image. We normalized each column of this map so that they have a unit median, and flagged all pixels with a resulting flux value larger than 5 as bad pixels. The optimal extraction was then repeated while forcing the masking of these previously identified bad pixels. 

The extracted spectra were normalized to a unit continuum and no wavelength calibration was performed at this step. The instrumental blaze function was also not corrected in this step, and is instead treated as a free parameter in our modeling of the science spectra when RVs are computed (see ${\S}$\ref{sec:pipeline}). This allows a better determination of the blaze function, especially in the presence of several deep absorption features from the gas cell.

\begin{figure}[p]
\centering
\includegraphics[width=0.8 \textwidth, clip=true,trim=0cm 0cm 0cm 0cm]{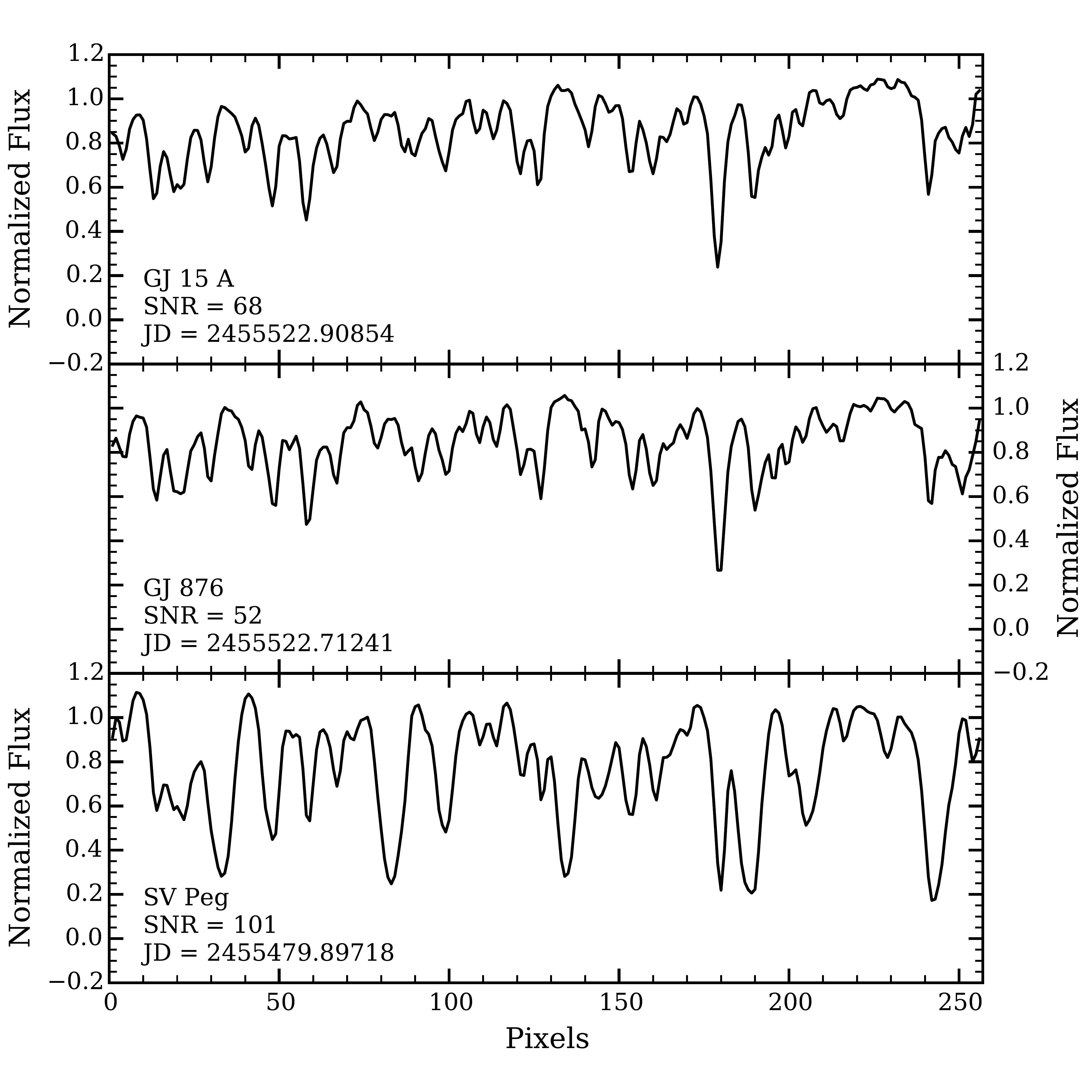}  
\caption{Sample observed spectra of GJ 15 A (top), GJ 876 (middle), and SV Peg (bottom) used in this work. The SNR per pixel and JD (2000) date of each spectra are shown at the bottom of each plot.}
\label{fig:spectra}
\end{figure}

A final bad pixel detection filter was applied in the extracted 1D spectrum by identifying up to 5 pixels with flux values significantly larger than the continuum. We chose this value of 5 pixels as it is the maximal number of bad pixels that we observed in an extracted 1D spectrum. We then counted the number of pixels with flux values $f_i > t$ where $t$ is a threshold value that goes from $1$ to $1.5$. If any range at least as large as $\Delta t = 0.05$ could be identified over which the number of flagged pixels did not change and was lower than 5, then these pixels were flagged as bad and ignored in the RV extraction pipeline. Figure \ref{fig:spectra} shows typical reduced spectra for each of the three targets. 

\section{Radial Velocity Pipeline}\label{sec:pipeline}

\begin{figure}[p]
\centering
\includegraphics[width=0.8\textwidth]{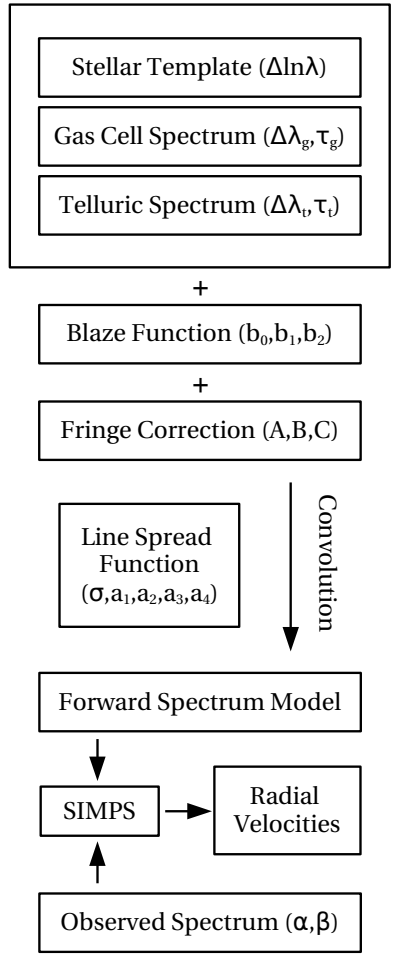}  
\caption{Schematic of the RV extraction pipeline.}
\label{fig:pipelineschematic}
\end{figure}

The RV extraction pipeline is written in Matlab with a few bookkeeping scripts written in IDL used for determining barycentric corrections (barycentric\textunderscore vel.pro, \citet[]{wright2014}). The RVs are retrieved by minimizing the difference between a model spectrum and the observed spectrum using the SIMPS Nelder-Mead amoeba simplex algorithm \citep[]{bajzer1999}. Figure \ref{fig:pipelineschematic} shows a schematic of the RV pipeline and includes all the free parameters that define the model spectrum. We elaborate on these parameters and the calculation of the RVs in the following sections\footnote{The RV pipeline described herein is currently private, but can be made available upon request to the authors. A future iteration of the pipeline dedicated to analyzing iSHELL data is planned to be made public once the code is sufficiently validated and documented.}.

\subsection{Spectral Model}\label{sec:spectralmodel}

We base the model $I_{obs}(\lambda)$ on that of \citet[]{butler1996}, where $\lambda$ is the wavelength solution of the model. However, additional components are necessary for our NIR observations, such that

\begin{equation}\label{eq:forwardmodel}
I_{obs}(\lambda) = LSF(\lambda) \ast [I_s(\lambda + \Delta \lambda_s) T_g(\lambda + \Delta \lambda_g) T_t(\lambda + \Delta \lambda_t)  \Sigma(\lambda) K(\lambda)],
\end{equation}

\noindent where $I_s$ is the stellar spectrum, which is derived iteratively (see ${\S}$\ref{sec:template}); $LSF$ is the line spread function/instrumental profile ($PSF$ in Eq. 1 of \citet[]{butler1996}); $\Sigma$ is a sinusoidal function that compensates for the interference fringing left in $I_{obs}$ by the CVF filter; $K$ is a quadratic blaze function that normalizes any curvature in the continuum, with the amplitude of each order ($b_0$, $b_1$, $b_2$) as free parameters; $\Delta \lambda_s$ and $\Delta \lambda_t$  are the Doppler shifts of the stellar and the telluric lines, respectively, and both are treated as free parameters; $T_t$ is the atmospheric telluric transmission function, with the line depths controlled by an exponent $\tau_t$

\begin{equation}\label{eq:taut}
T_t = T_{to}^{\tau_t} ,
\end{equation}

\noindent where $T_{to}$ is the NOAO telluric absorption spectrum given by \citet[]{livingston1991}; and $T_g$ is the methane isotopologue gas cell transmission function with its line depths controlled in an identical manner:

\begin{equation}\label{eq:taug}
T_g = T_{go}^{\tau_g}
\end{equation}

\noindent where $T_{go}$ was measured at high resolution using a Fourier Transform Spectrometer \citep[FTS;][]{plavchan2013}. $\tau_g$ is fixed at 0.97 under the assumption that the gas cell is stable over the observational period, with the specific value of 0.97 set through optimization tests. The fact that the fixed value is not 1.0 is likely due to the off-axis angle at which the cell was placed in the FTS, as opposed to in CSHELL, and the resulting slightly different light path lengths through the cell. $\tau_t$ is allowed to vary as a proxy for the airmass, though this is only valid in our case since most of the telluric lines in our wavelength range are from methane absorption; if water vapor absorption were also present, then their lines would behave differently, and a single $\tau_t$ would no longer suffice. 

\begin{figure}[p]
\centering
\includegraphics[width=0.8\textwidth]{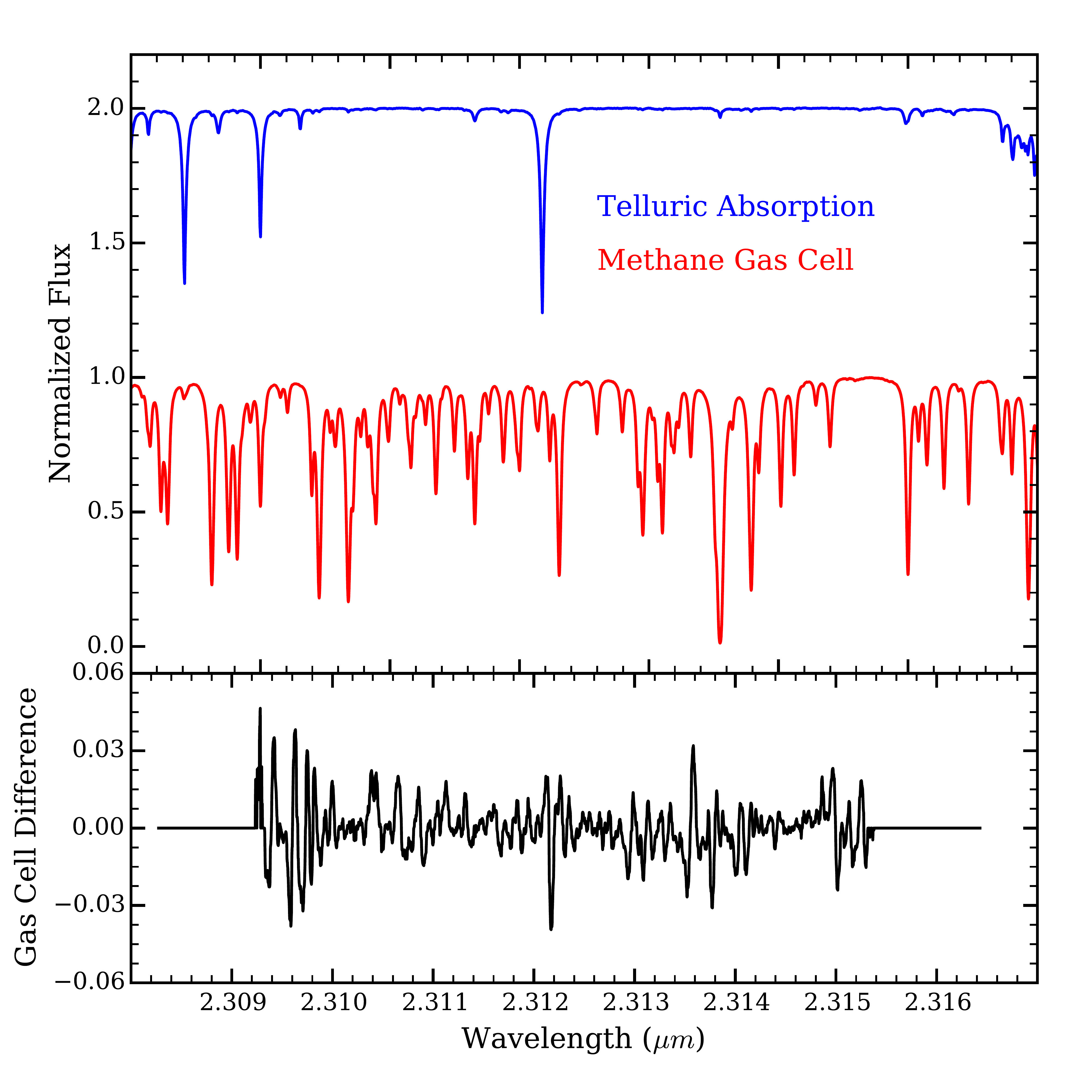}  
\caption{(Upper panel) Normalized telluric (blue, offset by +1) and methane isotopologue gas cell (red) absorption spectra used in the construction of our spectral model in the wavelength range of interest. (Lower panel) Difference between the laboratory gas cell spectra and the corrected gas cell spectra used in this work. }
\label{fig:tellgas}
\end{figure}

The upper panel of Figure \ref{fig:tellgas} shows $T_{go}$ (bottom) and $T_{to}$ (top) in the wavelength range of the observations. This range was chosen to minimize the number of contaminating telluric and OH emission lines while maximizing the number of gas cell lines used for wavelength calibration and the number of stellar CO lines to increase the RV information content of the target spectra. $T_{go}$ and $T_{to}$ have 12 and 5 times higher resolution than that of our observations, respectively. 

Even though $T_{go}$ is assumed to be much more accurately measured than  $T_{to}$ and $I_s$, we found features in the residuals of our fits of our spectral model to the observations that appeared to correspond to gas cell lines. To identify these features, we observed A type stars with the gas cell and fit to the data our spectral model using a flat line for $I_s$; the residuals were than averaged to reveal any coherent features, the majority of which should correspond to differences between the gas cell spectrum as observed at NASA IRTF and its laboratory-measured template, since the stellar spectrum has no features and there are only a few telluric lines. This process can be iterated multiple times by adding the averaged residuals to the gas cell and then repeating the fit and residual averaging. We expand on this process in ${\S}$\ref{sec:template}, where we use it to derive the stellar template. The lower panel of Figure \ref{fig:tellgas} shows the difference between the original $T_{go}$ and the final $T_{go}$ after multiple iterations. The lack of difference at the edges is due to the wavelength extent of the observations. There appears to be some correlation between the high frequency oscillations of the difference and the gas cell lines. The standard deviation of the difference is $\sim$ 1$\%$, and may be attributable to the differences in the environmental properties between where the gas cell template was measured and where it is being used for observations of our M dwarf targets, though the gas cell temperature is kept steady to within 0.1 K when used for observations \citep[]{plavchan2013}. Other possibilities include the existence of low amplitude, coherent noise in the detector that was not accounted for during the data reduction process, as well as subtle interpolation errors in the RV pipeline. Regardless of the cause, using the new $T_{go}$ reduces our RV scatter compared to using the original $T_{go}$.

$T_g$ is used to set the deviation of the gas cell spectrum, $\Delta{\lambda_g}$, from the ``true'' wavelength solution. In other words, we assume that the gas cell wavelength solution is correct, and any deviations of the gas cell lines in the observations from this true correct wavelength solution is due to a shift in the wavelength solution of the observations. We can then define a relative wavelength shift $\Delta{\lambda} = \Delta{\lambda_s} - \Delta{\lambda_g}$ that is directly related to the barycenter-corrected RV of the star with

\begin{equation}\label{eq:doppler}
RV = \frac{c \Delta \lambda}{\lambda_{c}} + v_b,
\end{equation}

\noindent where $c$ is the speed of light, $\lambda_{c}$ is the central wavelength of the spectral window, and $v_b$ is the barycenter velocity calculated using the IDL program barycentric\textunderscore vel.pro \citep[]{wright2014}. We perform the stellar wavelength shift in logarithmic space to be consistent with the Doppler equation, as $ \Delta{\ln{\lambda}} \sim \Delta \lambda / \lambda_{c}$. 

The LSF is constructed using Hermite functions $\psi_i(x)$ \citep[i.e.,][]{arfken2012}, which are derived iteratively using the following recursive relation and zeroth and first degree terms:

\begin{equation}\label{eq:hermiterecursion}
\psi_i(x) = \sqrt{\frac{2}{i}} [x\psi_{i-1}(x) - \sqrt{\frac{i-1}{2}} \psi_{i-2}(x)] ,
\end{equation}

\begin{equation}\label{eq:hermite0thterm}
\psi_0(x) = \pi^{-\frac{1}{4}} e^{-\frac{1}{2}x^2} ,
\end{equation}

\begin{equation}\label{eq:hermite1stterm}
\psi_1(x) = \sqrt{2}x\psi_0(x)  .
\end{equation}

\noindent where $x = P_f/\omega$, $P_f$ is the pixel grid of the model, and $\omega$ is the standard deviation (width) of the Gaussian factor, which is a free parameter in the model. We compute the LSF by summing the first $m$ terms

\begin{equation}\label{eq:lsf_definition}
LSF(x) = \psi_0(x) + \sum \limits_{i = 1}^m a_i \psi_i(x)
\end{equation}

\noindent where $a_i$ is the amplitude of the ith term relative to the 0th term and are also free parameters. Our nominal model uses $m$ = 4, which we show to be sufficient in ${\S}$\ref{sec:gj15a_lsf_blaze}. The LSF is normalized to a unit area under curve, and the simplex minimization is constrained such that no part of the LSF becomes negative. We experimented with multi-Gaussian and multi-Lorentzian LSFs, as in previous works \citep[e.g.,][]{valenti1995,butler1996,johnson2006,bean2007,bean2010}, but they yielded lower RV precision than our current Hermite function implementation, likely as a result of the relatively low SNR of our observations. The convolution is done on the pixel grid rather than the wavelength grid, though they are equivalent. 

\begin{figure}[p]
\centering
\includegraphics[width=0.8\textwidth]{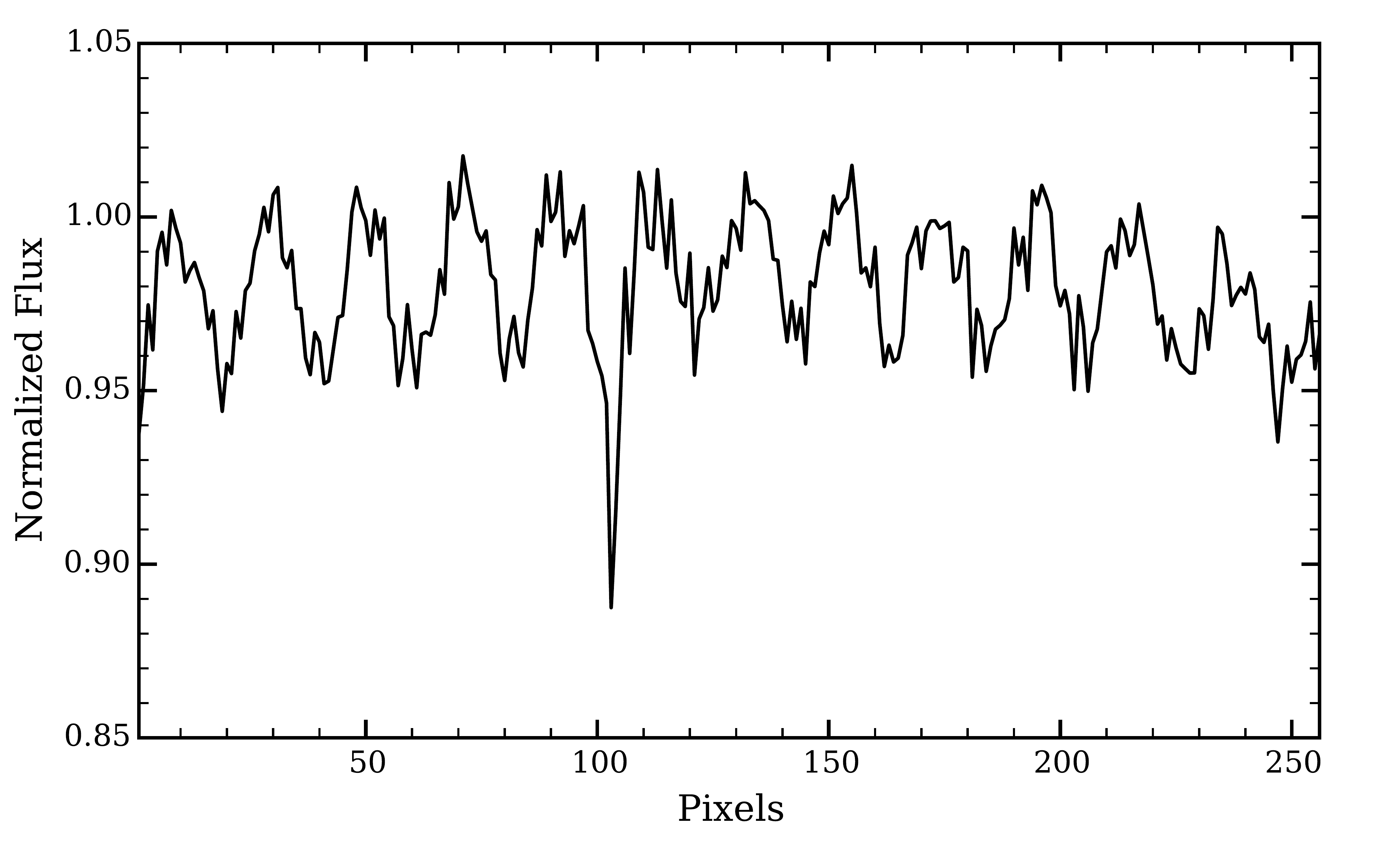}  
\caption{H band spectrum of 32 Peg taken without the methane isotopologue gas cell, showing the prominent sinusoidal fringing as a result of the CSHELL CVF filter. Similar fringing is found in all of our K band data. The absorption line near pixel 100 is an atmospheric telluric feature.}
\label{fig:astarsinusoid}
\end{figure}

As described in ${\S}$\ref{sec:reduction}, sinusoidal interference patterns of variable amplitude and phase are present in both the flat fields and the science data due to the CSHELL CVF filter. These fringes become obvious in the spectrum of an A star taken without the methane isotopologue gas cell, as shown for 32 Peg in Figure \ref{fig:astarsinusoid}. Fringes of a similar amplitude were also seen in the spectra taken by \citet[]{brown2002} using NIRSPEC, and were corrected by parametrization of the sinusoid and linear regression. Meanwhile, \citet[]{blake2010} also observed fringes while using NIRSPEC, which they corrected by applying a model sinusoid of fixed amplitude, period, and phase. We use a similar method and parametrize the fringing as a sinusoid function $\Sigma(\lambda)$:

\begin{equation}\label{eq:sinusoid}
\Sigma = 1 + A \sin{\left(\frac{2 \pi \lambda}{B}  + C\right)} ,
\end{equation}

\noindent where the parameters $A$, $B$, and $C$ are free to vary.

The spectral model is compared to the data on a common wavelength grid. As with \citet[]{crockett2011}, we find that the wavelength solution of the data is variable between observations, and that the variability is quadratic in wavelength. We also find that it is dependent on the trace position $y$ of the observed spectrum on the detector. Thus, we define the wavelength solution of the observations, $\lambda_o$, with respect to $\lambda$ by

\begin{equation}\label{eq:wavelengthscale}
\lambda_{o} = \lambda  + \alpha' \left(\frac{P_o - P_o^c}{N_{pix}}\right) + \beta' \left(\frac{P_o - P_o^c}{N_{pix}}\right)^2 ,
\end{equation}

\noindent where $\alpha'$ is a linear function of $y$ and $\beta'$ is constant in $y$

\begin{equation}\label{eq:alphaypos}
\alpha' = -0.58 + 0.0051y + \alpha
\end{equation}

\begin{equation}\label{eq:betaypos}
\beta' = -0.4 + \beta
\end{equation}

\noindent $\alpha$ and $\beta$ are free parameters, $P_o$ is again the pixel grid on the detector, $P^c_o$ is the central pixel of said pixel grid, and $N_{pix}$ = 256 is the total number of pixels in the data. The coefficients of Eqs. \ref{eq:alphaypos} and \ref{eq:betaypos} are derived from fitting a linear trend to the relationship between $y$ and a set of $\alpha'$ and $\beta'$ values obtained from fits to several high SNR observations of GJ 537 A \citep[]{gagne2015} where no $y$ dependence was assumed (i.e. $\alpha' = \alpha$, $\beta' = \beta$)

The model $I_{obs}(\lambda)$ is then interpolated onto $\lambda_{o}$. We use 4096 pixels in the model pixel grid $P_f$ to reach sufficient resolution for convolution with the LSF, and downsample to 256 pixels when interpolating onto $\lambda_{d}$. The downsampling is done by binning every $4096/256 = 16$ model ``pixels''. Any model pixel that is lying partly in two data pixel bins is split between them according to the fraction of the model pixel that is in each of the two data pixel bins.

Table \ref{table:rvparams} lists all of the free parameters that define the nominal spectral model, which are varied by SIMPS to optimize the fit between the model and the observed spectra. The optimization is accomplished by minimizing the average of the RMS and the robust sigma \citep[]{hoaglin1983,beers1990} of the residuals after application of the bad pixel mask, which sets the weight of bad pixels to zero (see ${\S}$\ref{sec:reduction}). Including the robust sigma in the minimization process decreases the impact of outliers due to noise, hot pixels, and unmasked bad pixels. However, just minimizing the robust sigma alone without the RMS produces bad fits to the telluric lines due to their similarity to outliers when there are very few of them, as in our case (Figure \ref{fig:tellgas}). All parameters aside from the gas cell optical depth $\tau_g$ are allowed to vary, while only the relative wavelength shift $\Delta{\ln{\lambda}}$, the gas cell wavelength shift $\Delta{\lambda_g}$, and the phase of the fringing correction $C$ are allowed to vary freely. The other parameters are allowed to vary within bounds determined from optimization tests. 

\subsection{RV Calculation}\label{sec:rvcalc}

Due to the low SNR per pixel (Table \ref{table:observations}) and small wavelength range ($\sim$ 6 nm) of our individual spectra, the RV precision of a target is calculated from nightly averaged RV values $RV_i$, for the ith night, defined as the weighted mean of the RVs of all the individual spectra taken during that night, weighted by the inverse square of the RMS of residuals of each model fit to those spectra, ignoring the contributions to the RMS from bad pixels,

\begin{equation}\label{eq:nightlyrv}
RV_i = \frac{\sum \limits_{j} w^j_i RV^j_i}{\sum \limits_{j} w^j_i} 
\end{equation}

\begin{equation}\label{eq:nightlyrvweight}
w^j_i =  \left\{ \frac{1}{N_{pix}-1} \sum \limits_{k \neq k_{bp}}^{N_{pix}} [I^j_{d,i}(k)  - I^j_{obs,i}(k)]^2 \right\} ^{-1}
\end{equation}

\noindent where $RV^j_i$, $w^j_i$, $I^j_{d,i}(k)$, and $I^j_{obs,i}(k)$ are the RV, weight, observed flux of the kth pixel, and model flux of the kth pixel of the jth individual spectra on the ith night, respectively, with $k_{bp}$ indicating a bad pixel. The $1\sigma$ error bar of the ith nightly averaged RV point, $\delta RV_i$, is calculated as the weighted standard deviation of the individual RVs, weighted by the same quantity as calculated for the nightly averaged RV values (see Eq. \ref{eq:nightlyrvweight}), and divided by the square root of the number of spectra $N_{obs}$ taken during that night,

\begin{equation}\label{eq:nightlyrvweightstd}
\delta RV_i =  \left\{ \frac{\sum \limits_{j} [w^j_i (RV^j_i - RV_i)]^2}{N_{obs} \sum \limits_{j} w^j_i} \right\} ^{\frac{1}{2}}
\end{equation}

\noindent The RV precision of a target is then defined as the standard deviation of the nightly averaged RV values, while the reduced chi-square $\chi^2_{red}$ is calculated using 

\begin{equation}\label{eq:redchisqrd}
\chi^2_{red} = \frac{1}{N_n - 1} \sum \limits_i^{N_n} \frac{(RV_i - \overline{RV})^2}{\delta RV_i^2}
\end{equation}

\noindent where $N_n$ is the number of nightly averaged RV points (= number of nights of observations/epochs), and $\overline{RV}$ is the weighted mean of all the $RV_i$ values, weighted by the inverse square of the $\delta RV_i$ values. 

Note that our method of calculating the RV precision of a target is different from that of \citet[]{gagne2015}, our companion survey paper. They define the RV precision as the weighted standard deviation of the nightly averaged RV values, weighted by $\delta RV_i$, which allows them to better constrain planet mass sensitivity. In contrast, the purpose of the RV precision obtained in our work is to compare to the RV precision obtained by previous works, where it is sufficient to calculate just the standard deviation of the nightly averaged RV values. The method of \citet[]{gagne2015} results in lower RV RMS overall, which is indicative of mixing data of different SNRs. 


\subsection{Stellar Template Generation}\label{sec:template}

The retrieval of the original stellar spectrum, $I_s$, for use in Eq. \ref{eq:forwardmodel} has consistently been a difficult task. Some previous works have used synthetic stellar models that calculated stellar spectra, given effective temperatures and surface gravities \citep[]{blake2010,crockett2011,bailey2012,tanner2012}, but this runs the risk of introducing spectral features not present in the science targets into the spectral fitting, which lowers the RV precision. Alternatively, the stellar spectrum can be obtained from deconvolution of high resolution stellar observations using LSFs derived from observations of A and B stars taken with a gas cell \citep[]{butler1996,bean2010,rodler2012}. However, this presumes that the LSF remains stable between the observations of the targets and those of the A and B stars, which is not true for our case due to temperature variations, instrumental flexure, and mechanical disturbances to CSHELL itself (e.g. moving the slit into position) as it moves with the telescope at its Cassegrain mount. 

We thus use the target observations themselves to derive the stellar template iteratively using a method similar to that of \citet[]{sato2002}. However, our method differs from theirs in that the initial guess template is completely flat in order to minimize contamination of features that may not be present in the observed spectra, and that we use all of our target observations to derive the template in order to maximize SNR and sample a large range in barycenter velocities. This ensures that the stellar spectrum is decoupled from ``stationary'' features, such as the gas cell and telluric lines and the sinusoidal fringing, so that they do not contaminate the stellar template. 

The procedure begins with the building of a model using the flat spectrum in place of $I_s$ in Eq. \ref{eq:forwardmodel}. The fit produces residuals similar in shape to the actual stellar spectrum, though they are distorted due to the simplex algorithm minimizing the residuals; for example, the stellar CO lines will be made more shallow, or raised above the continuum to reduce their impact on the RMS and robust sigma. We lessen this effect by repeating the first iteration with the CO lines masked out, allowing only the stellar continuum to contribute to the RMS and robust sigma. The CO lines are masked by ignoring all pixels with values one standard deviation below the mean of the normalized template flux derived from the original first iteration; two pixels on each side of each CO line are also masked to further reduce their impact on the fit. 

The residuals are then ``de-shifted'' to the barycenter so that they all have the same Doppler shift, with the assumption that the star has zero RV perturbations. This is valid even for potential planet hosts, as RV perturbations caused by planets are much smaller than a single resolution element. Residual values that are three standard deviations above or six standard deviations below the mean residual value (zero), and those below $-1$, are set to zero so that they do not contribute to the eventual template (i.e. they are assumed to be bad pixels). We then take the median of these residuals, weighted by the inverse square of the average of the RMS and robust sigma of each fit, and add it to the template from the previous iteration to generate the new template that will be used in the next iteration. Pixels that are flagged by the bad pixel mask are given zero weight. In other words, residuals from better fits are more represented in the stellar template than those from worse fits. We find that variations in the quality of fit is usually due to the quality (SNR) of the data itself, and therefore it is reasonable to rely more on the best data to generate the stellar template. 

This process is repeated until the RV precision of the target stabilizes, which usually takes $\sim$ 10 iterations, though we typically run the pipeline for 20 iterations to confirm that RV stabilization has been reached. We speed up this process by deconvolving the residuals from the first iteration with the best fit LSF corresponding to each residual using the Richardson-Lucy algorithm \citep[]{richardson1972,lucy1974}, so as to more quickly converge to the ``real shape'' of the stellar spectrum. Deconvolution is avoided in subsequent iterations, as it would significantly amplify the noise in the data. 

\section{Results and Discussion}\label{sec:results}

\subsection{RV Stability at Modest Signal-to-Noise: GJ 15 A}\label{sec:gj15a_results}

\subsubsection{Radial Velocities}\label{sec:gj15a_rvs}

\begin{figure}[p]
\centering
\includegraphics[width=0.8\textwidth]{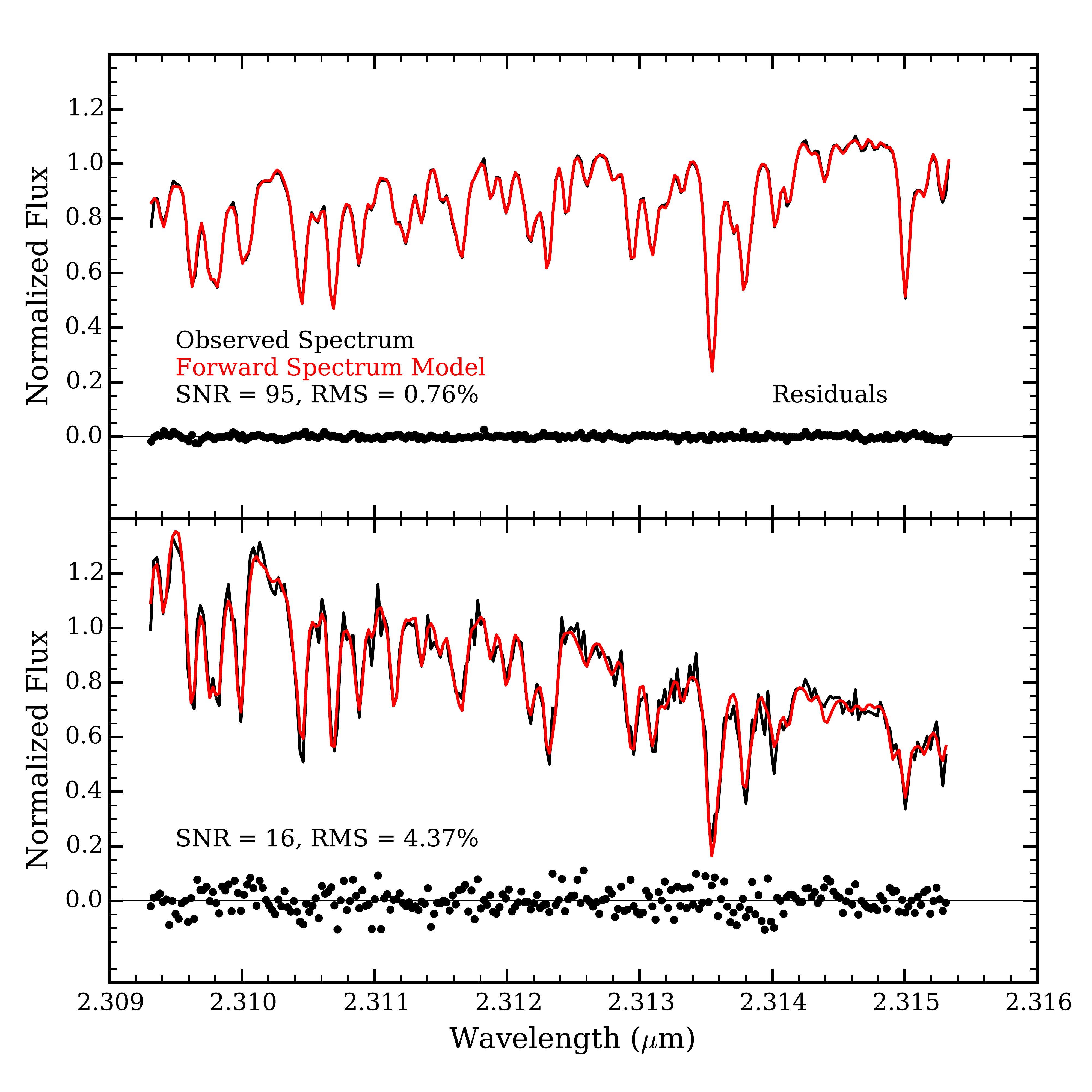}  
\caption{Examples of optimized fits to GJ 15 A spectra. (Top) Fitting a spectral model (red) to a high SNR observation (black line), with resulting residuals (black points). (Bottom) Fitting a spectral model to a low SNR observation. }
\label{fig:gj15a_fitresults}
\end{figure}

Figure \ref{fig:gj15a_fitresults} shows example fits of the model spectrum to a high SNR per pixel (top) and a low SNR per pixel (bottom) observed spectrum of GJ 15 A. Note that the high SNR spectrum is not included in our RV calculations, as it was the only spectrum taken that night. It is clear that, at SNR $\sim$ 100, our spectral model is able to reproduce the observations with high fidelity, while at low SNR there is significant scatter and deviation between model and data, though all of the major features have been captured by the model. As Table \ref{table:observations} shows, the majority of our observations have SNR per pixel closer to $\sim$ 30\textendash40, thus limiting our RV precision. We will quantify the effect of the SNR on our achievable RV stability in ${\S}$\ref{sec:gj15a_err}.

\begin{figure}[p]
\centering
\includegraphics[width=0.8\textwidth]{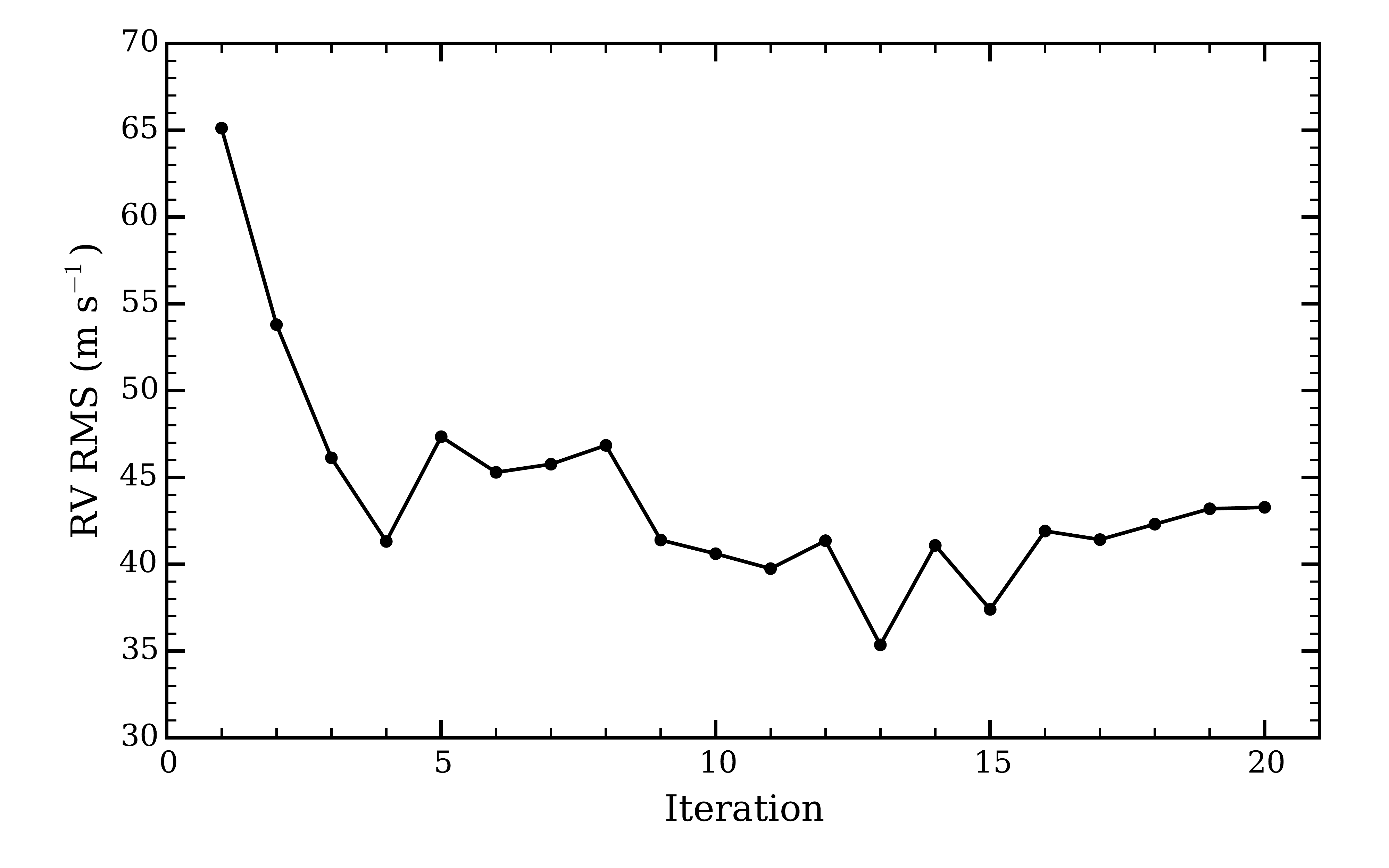}  
\caption{GJ 15 A RV RMS, defined as the standard deviation of the nightly RV points, as a function of iterations of the RV extraction pipeline (points).}
\label{fig:gj15a_fitrms_vs_iteration}
\end{figure}

\begin{figure}[p]
\centering
\includegraphics[width=0.8\textwidth]{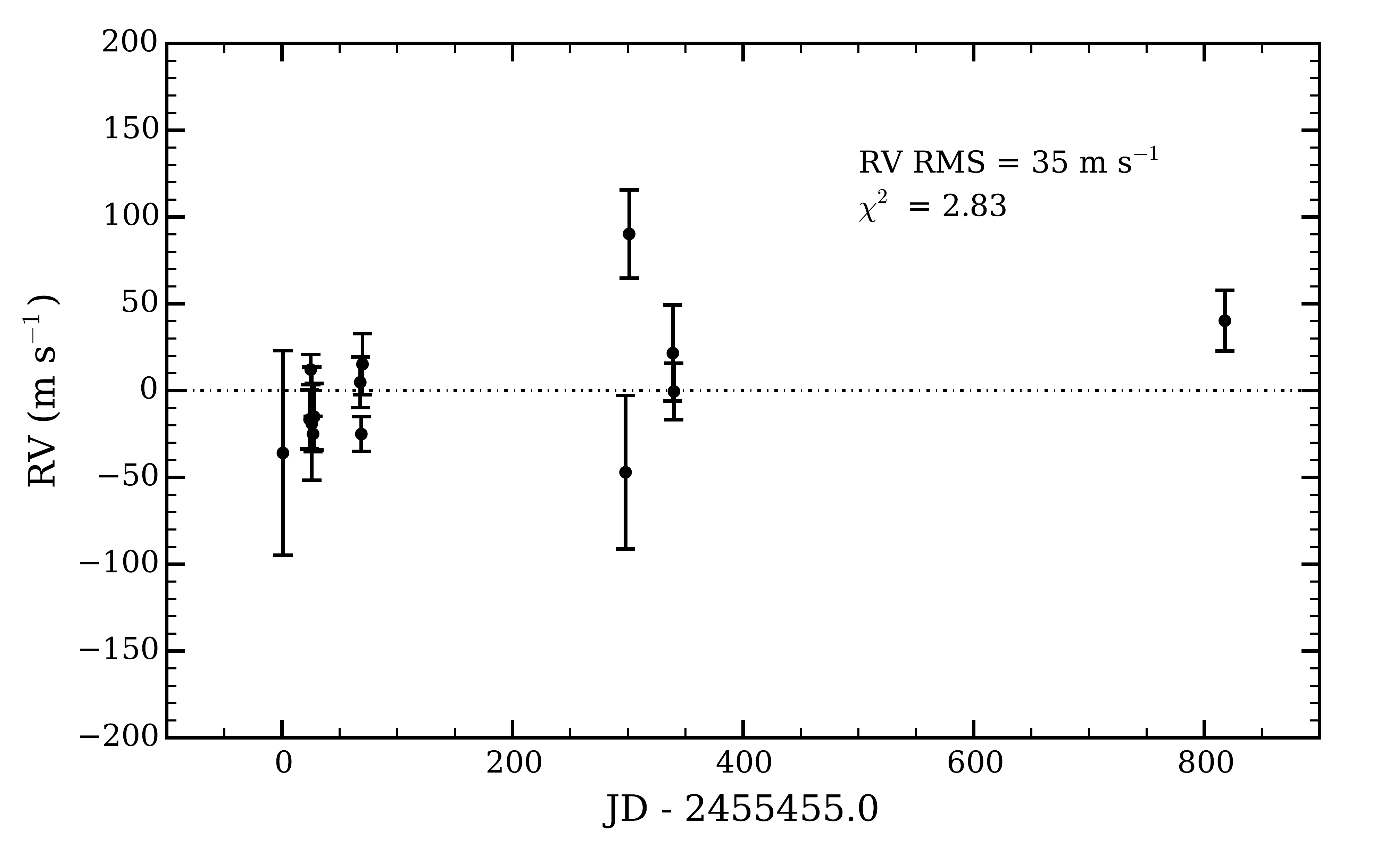}  
\caption{The nightly RVs of GJ 15 A from the iteration of the RV extraction pipeline with the lowest RV RMS.}
\label{fig:gj15a_rv}
\end{figure}

The iterative nature of our RV pipeline results in multiple values of RV precision for a single target, one for each iteration, and we accept the lowest RV scatter (RMS) value among all the iterations as the RV precision achievable by our RV pipeline for that target. Figure \ref{fig:gj15a_fitrms_vs_iteration} shows the RV scatter as a function of iteration for GJ 15 A. The first few iterations with high RV scatter result from errors in the stellar template, as it is continually augmented from one iteration to the next. Convergence in both the stellar template and the RVs is reached beyond iteration 9, after which the RV scatter is $\sim$ 40 $\ms$, with occasional deviations to lower values. Figure \ref{fig:gj15a_rv} shows the RVs from iteration 13, where we are able to achieve a RV precision of 35 $\ms$ over a 817 day long baseline. Table \ref{table:gj15a_gj876_svpeg_rv} gives the nightly RV values and associated $1\sigma$ error bars. The obtained RV precision is consistent with the theoretical RV precision given our observational setup and a SNR per pixel of $\sim$ 100, the effective SNR of each of our nightly averaged RV points \citep[see Table 2 of][]{anglada2012}. If the individual RV points are considered rather than the nightly RVs, then the RV precision is 131 $\ms$, and the SNR of the observations are those between the minimum and maximum SNR values listed in Table \ref{table:observations}. 

\begin{figure}[p]
\centering
\includegraphics[width=0.5\textwidth]{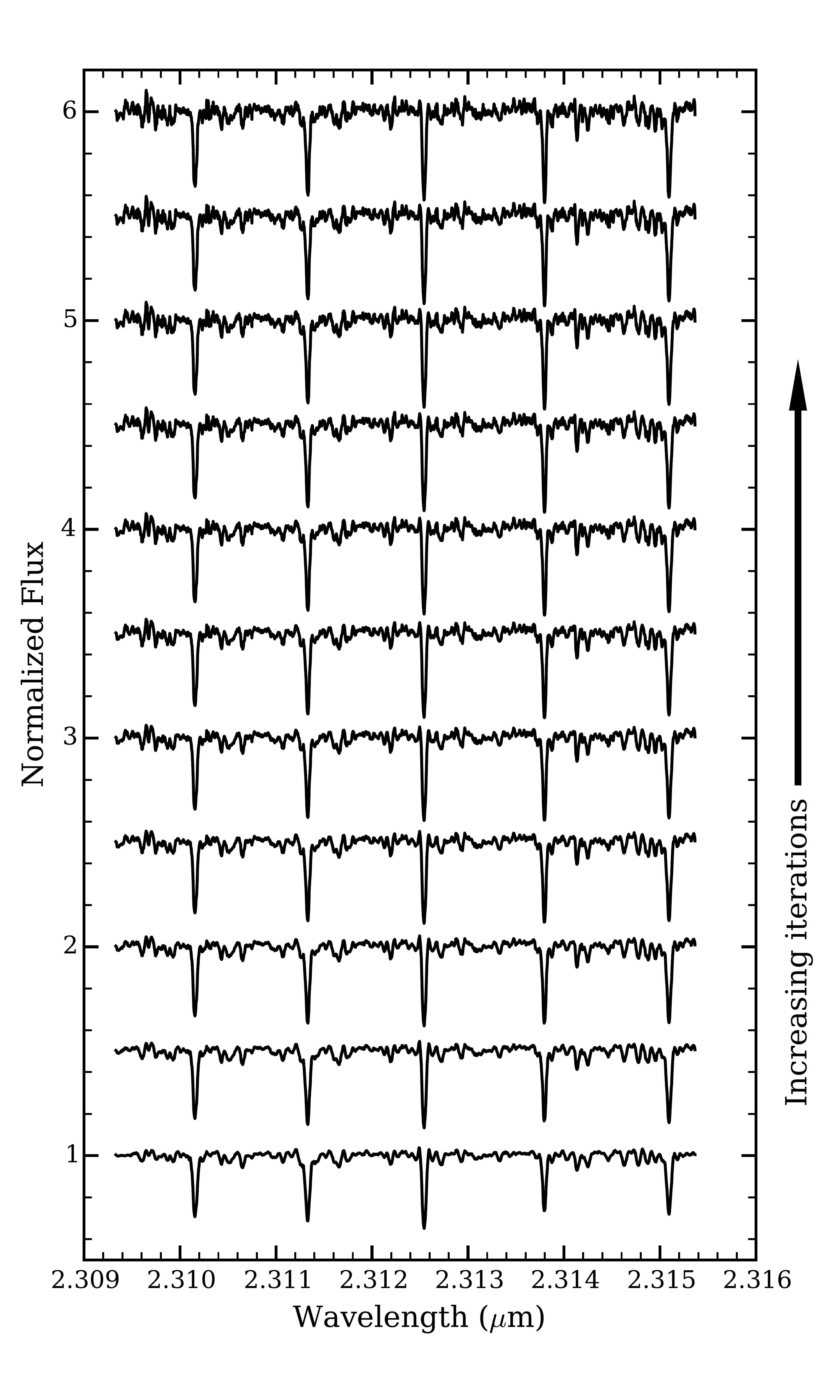}  
\caption{Evolution of the GJ 15 A stellar template with increasing iterations (upwards) of the RV extraction pipeline. All templates aside from the first (bottom) are shifted upwards by 0.5 to avoid overlap.}
\label{fig:gj15a_template_iteration}
\end{figure}

Figure \ref{fig:gj15a_template_iteration} shows the progression of the stellar template with increasing iteration. In the initial iterations, the depth of several of the CO lines are much shallower than the central CO line, when they should be similar in depth instead. This is due to some overlap between these stellar lines and the gas cell lines, allowing the gas cell template to partially compensate for them during the spectral fits. However, at higher iterations, improved fits to the observations arising from improved stellar templates lead to similar CO line depths. The higher iterations also show more noise in the templates, which is a drawback of this algorithm, as any coherent noise in the residuals will be added to the template. Therefore, at higher iterations continuous augmentation of the stellar template does not lead to higher RV precision, as shown in Figure \ref{fig:gj15a_fitrms_vs_iteration}.

\subsubsection{Parameter Correlations}\label{sec:gj15a_corr}

Given our large number of nuisance parameters (Table \ref{table:rvparams}), it is important to investigate any correlations between themselves and between them and the RVs. Correlations are tested by evaluating Pearson's linear correlation coefficient $\rho$, defined by \citep[]{pearson1895}

\begin{equation}\label{eq:pearson}
\rho = \frac{cov(X,Y)}{\sigma_X \sigma_Y} , 
\end{equation}

where $cov(X,Y)$ is the covariance of the variables $X$ and $Y$, and $\sigma_X$ and $\sigma_Y$ are their standard deviations. A large degree of correlation corresponds to $|\rho| \rightarrow 1$. We calculate $\rho$ by assigning the RVs and different spectral model parameters to $X$ and $Y$. Figure \ref{fig:gj15a_rv_vs_param} shows the relationship between every parameter and the individual RVs of GJ 15 A. No obvious trends can be discerned for any of the parameters, while the absolute value of $\rho$ is $<0.4$ for all of the RV\textendash parameter pairings. Some parameters have hit their parameter bounds, such as the parameters controlling the shape of the LSF. For example, the amplitude of the first degree Hermite function (a$_1$) hits the bounds in both directions; this is understandable since it shifts the LSF back and forth, which could create a false RV signal, and thus it is necessary to constrain it to a narrow set of values. The FWHM of the LSF (2$\omega \sqrt{2Ln2}$) spans the critical sampling resolution at the precision of CSHELL ($\sim$ 2 pixels/resolution element), and therefore we are both under and oversampling our science spectra. Clustering of points can be seen in some of the panels. For $\Delta{\ln{\lambda}}$, the clusters correspond to different epochs with different barycentric corrections (see Eq. \ref{eq:doppler} and the discussions that follow). For $\Delta{\lambda_g}$ and $\Delta{\lambda_t}$, the clustering corresponds to different epochs where the central wavelengths were set to slightly different values. Variations in the LSF parameters are due to the variability inherent in observing using a non-stabilized spectrograph. Figure \ref{fig:gj15a_lsfvar} shows the variability in the LSF within a single night (top) and between several nights (bottom). To quantify the variability, we take the average of the standard deviation of the difference between the LSFs and their mean within 4 pixels of the x-grid zero point. The resulting $\sigma_{LSF}$ values show that the LSF is about 4 times more variable between nights than within a night, which could account for some of the systematic RV differences between nights. 

While no parameters show unexplained increasing or decreasing trends with RV, several parameters do have ranges in values where the corresponding RV scatter is greater than that of other values of said parameters. For example, low values of the sinusoidal fringing amplitude ($A$) appear to correspond to RV points with lower scatter than those that correspond to high values of $A$. Other examples of this phenomenon include the blaze function polynomial coefficients (b$_0$, b$_1$, b$_2$) and the wavelength solution parameters ($\alpha$, $\beta$). These trends in RV scatter are shown more clearly in Figure \ref{fig:gj15a_binned_params}, where we bin every 10 RV points in the order of increasing value of the chosen parameters and calculate their RV scatter. We thus see that RV precision appears to increase with increasing continuum level, decreasing (in magnitude) linear and quadratic terms in the blaze function, decreasing sinusoidal fringing amplitude, and decreasing (in magnitude) linear and quadratic terms in the wavelength solution.  

The absolute value of $\rho$ is $<0.5$ for most of the parameter\textendash parameter pairings. The largest $|\rho|$ (0.9985) is obtained between the gas cell wavelength shift $\Delta{\lambda_g}$ and the telluric wavelength shift $\Delta{\lambda_t}$, which is expected since they both act as wavelength calibration, even though the gas cell is a more reliable fiducial. Aside from this correlation, the largest $|\rho|$ values ($\geqslant$0.6) occur between the six parameters shown in Figure \ref{fig:gj15a_binned_params}, and their correlations and $\rho$ values are shown in Figure \ref{fig:gj15a_paramcorr}.

For the correlations between the linear and quadratic terms of the blaze function ($b_1$ and $b_2$) and wavelength solution ($\alpha$ and $\beta$), anti-correlations may arise due to each pair being alternating orders in a polynomial. In addition, the change in shape of the continuum slope may play a role as well. Consider a linear slope in the continuum of a spectrum;  if the data wavelength solution is expanded in both directions equally, corresponding to a more positive $\alpha$, then the linear slope of the data will tend towards zero, which, since most of the linear slopes in the sample are negative, results in a more positive model linear slope $b_1$. This explains the positive correlation (large positive $\rho$) between $\alpha$ and $b_1$. Similarly, a positive $\beta$ corresponds to an expansion at longer wavelengths and compression at shorter wavelengths; this results in the generation of positive curvature when the linear slope is negative, corresponding to a positive $b_2$, and thus the positive correlation between $b_2$ and $\beta$. We will see in the next section that {some of the remaining correlations, such as those between the constant blaze function term $b_0$ and the fringing amplitude $A$} are related to the SNR of the spectra.

\subsubsection{Error Analysis}\label{sec:gj15a_err}

Typical contributions to the RV scatter of a target include photon noise, wavelength calibration errors, instrumental effects, and astrophysical sources, such as stellar activity \citep[]{bailey2012}. GJ 15 A has been shown to be stable down to 3.2 $\ms$ \citep[]{howard2014}, with the dispersion dominated by the RV signal from a Super-Earth exoplanet (K = 2.94 $\ms$); this can be regarded as the total error contribution from astrophysical sources, and it is clearly far lower than our calculated RV precision. Therefore, our RV error budget is dominated by the other terms. 

We now calculate the error contribution from photon noise, $\sigma_{RV}$, due to the RV information content of the stellar spectrum and the SNR of the observations. Following \citet[]{butler1996}, 

\begin{equation}\label{eq:sigmarv}
\sigma_{RV} = \left[  \sum \limits_{i} \left ( \frac{dI_i / dV_i}{\epsilon_i} \right )^2  \right]^{-\frac{1}{2}} , 
\end{equation}

\noindent where $dI_i/dV_i$ is the slope of the stellar spectrum at the ith pixel, calculated as the change in the normalized flux of the spectrum at the ith pixel, $I_i$, divided by the corresponding change in wavelength, expressed in units of velocity, at the ith pixel; $\epsilon_i$ is the uncertainty in the flux at the ith pixel, given by

\begin{equation}\label{eq:epsiloni}
\epsilon_i = \frac{1}{\sqrt{N_{pho}}} = \frac{1}{SNR\sqrt{I_i}}
\end{equation}

\noindent in the photon-limited case, where $N_{pho}$ is the number of photons and the SNR is that of the observation (Table \ref{table:observations}). 

Due to our much smaller wavelength range compared to that of \citet[]{butler1996}, we use Eqs. \ref{eq:sigmarv}\textendash \ref{eq:epsiloni} differently. In their work, they split their one observed stellar spectrum of $\tau$ Ceti into 704 2 \AA${}$ segments and calculated $\sigma_{RV}$ for each segment (i.e. the summation in Eq. \ref{eq:sigmarv} is over the pixels of a single segment); they then compiled a histogram of $\sigma_{RV}$ values for the segments, and calculated the RV precision arising from combining all the segments. Our small wavelength range forbids us from splitting our observations into multiple segments, so instead we treat each observation as a single ``segment'', giving us 103 segment in total for GJ 15 A; we then calculate $\sigma_{RV}$ for each observation/segment by summing over every pixel in the observation in Eq. \ref{eq:sigmarv} and using the SNR per pixel of each observation in Eq. \ref{eq:epsiloni}.

An additional complication is that we do not have any observations of the star by itself, as even observations taken without the gas cell are contaminated by atmospheric telluric lines. Therefore, we build a synthetic stellar spectrum for each observation using the stellar template for the iteration with the best RV precision. The synthetic spectrum is calculated by (1) downsampling the stellar template to the data wavelength solution derived from the spectral fit of the model to that particular observation, (2) convolving the result with the LSF derived from the same fit, and (3) adding onto the result the residuals from the fit to simulate the appropriate level of noise. 

The upper panel of Figure \ref{fig:gj15a_rverr} gives a histogram of the calculated $\sigma_{RV}$ values of the 103 individual observations taken for GJ 15 A. The most common $\sigma_{RV}$ value is $\sim$ 120 $\ms$, slightly lower than 131 $\ms$, the actual RV precision for the individual observations (not the nightly coadded RV points) as output by our RV pipeline. 

We can repeat steps (1) and (2) above for the gas cell spectrum to find the RV scatter contribution from errors in the wavelength calibration. We omit step (3) as the gas cell does not contribute to the photon noise. The lower panel of Figure \ref{fig:gj15a_rverr} shows the result. Given the higher line density of the gas cell spectrum it is not surprising that its $\sigma_{RV}$ values are much lower than that of the stellar spectrum. The peak $\sigma_{RV}$ value is $\sim$ 50 $\ms$. Combining this with the photon-limited errors calculated above in quadrature gives $\sqrt{120^2 + 50^2}$ = 130 $\ms$, essentially equal to the RV precision of the actual individual observations. We can thus conclude that our RV scatter is dominated by photon noise, and to a lesser extent from errors in wavelength calibration.  
 
We can also estimate the theoretical nightly averaged RV scatter by assuming the 103 individual observations are spread out evenly among the 14 nights, such that the RV scatter should decrease by $\sqrt{103/14} \sim 2.7$, so that it equals 131/2.7 $\sim$ 48 $\ms$, slightly higher than  our actual results (Figure \ref{fig:gj15a_fitrms_vs_iteration}). This is likely due to the fact that the observations are not spread out evenly among the nights.  

Figure \ref{fig:gj15a_rverr} shows a small local maximum at higher values of $\sigma_{RV}$, which correspond to the low SNR observations of the night of July 13th, 2011 (Table \ref{table:observations}). As the RV content and wavelength calibration of each observation are nearly identical, the chief factor causing the higher $\sigma_{RV}$ values is the low SNR. The upper panel of Figure \ref{fig:gj15a_rv_vs_snr} shows the relationship between RV and SNR per pixel. There is a clear trend showing that RVs of low SNR observations have greater scatter than RVs of high SNR observations. We can thus estimate the impact the SNR has on the RV precision by repeating the calculations of Figure \ref{fig:gj15a_binned_params}, but with the parameter values replaced by SNR values; the result of this calculation is shown in the lower panel of Figure \ref{fig:gj15a_rv_vs_snr}. To quantify the relationship between RV scatter and SNR, we fit a power law to the points with the form

\begin{equation}\label{eq:sigmarvsnr}
\sigma_{RV}^2 = \sigma_0^2 + \left (s_1 SNR \right )^{s_2}, 
\end{equation}

\noindent where $\sigma_0$ is the error contribution from all other sources and is a free parameter in the fit, and $s_1$ and $s_2$ are constants that are also free parameters. The power law is consistent with a 1/SNR relationship between $\sigma_{RV}$ and SNR, as expected when the main source of error is due to photon noise. 

Given the impact the SNR has on the RV scatter, it is useful to investigate how SNR variations affect the free parameters in the spectral model, which is shown in Figure \ref{fig:gj15a_params_vs_snr}. As with Figure \ref{fig:gj15a_rv_vs_param}, most of the parameters show no clear trends with SNR, except for the blaze function coefficients, fringing sinusoid amplitude, and the linear and quadratic terms in the wavelength solution. It is understandable why these parameters are the most affected. At low SNR, neither the continuum nor the wavelength solution are well constrained due to the noise in the spectra, leading to large deviations from zero (or 1.05, in the case of the constant term in the blaze function). The rise in fringing amplitude with decreasing SNR can be seen as either an actual increase in the relative amplitude of the fringing as the number of photons from the star decreases, or as the model attempting to fit the higher noise level with a sinusoid in low SNR spectra.

These results help to explain some of the correlations between the parameters seen in Figure \ref{fig:gj15a_paramcorr}. For example, as the blaze function constant term $b_0$ can only decrease and the fringing amplitude $A$ can only increase at low SNR, they appear to correlate with each other, whereas the true cause of the correlation is the low SNR. Similar arguments can be made for correlations between these two parameters and the other four parameters of Figure \ref{fig:gj15a_paramcorr}.

Sources of error stemming from instrumental effects are likely dominated by contributions from bad pixels, which were abundant in the science images and can be generally defined as due to detector imperfections and cosmic rays. Figure \ref{fig:gj15a_rv_vs_bp} shows the relationship between RV scatter and the number of bad pixels N$_{bp}$ flagged on the 2D trace of GJ 15 A spectra. As with Figure \ref{fig:gj15a_rv_vs_snr}, the upper panel plots the individual RV values against N$_{bp}$, while the lower panel shows the RV scatter for every 10 points binned in increasing N$_{bp}$. We once again fit a power law to the binned points, though we ignore all points corresponding to N$_{bp}$ $>$ 150 since our knowledge about the RV precision for this range of N$_{bp}$ is limited by the small number of spectra. The fitted curve is the same as in Eq. \ref{eq:sigmarvsnr}, with N$_{bp}$ in place of the SNR. The initial steep rise in $\sigma_{RV}$ from N$_{bp}$ = 0 to N$_{bp}$ = 150 shows the significant impact bad pixels have on both the data reduction and retrieval of RVs. Beyond N$_{bp}$ = 150, the RV scatter appears to saturate. Bad pixels not only cause certain parts of the spectrum to be unusable, thereby decreasing the information content of the data, but they also affect the data reduction process by interfering with the initial fitting to the spatial profile of the 2D trace.

\subsubsection{Exploring the Multiplicity of LSF and Blaze Function Terms}\label{sec:gj15a_lsf_blaze}

The LSF and blaze function components of our spectral model are constructed from the summation of multiple Hermite function and polynomial terms, respectively. However, there are no significant \textit{a priori} constraints on how many terms these components should have. We thus perform sensitivity tests by running our RV pipeline using data from GJ 15 A and various numbers of component terms in the LSF and blaze function and then comparing the resulting RV precision.  

Figure \ref{fig:gj15a_lsf_test} shows the resulting RV precision when using LSFs constructed with up to 11 Hermite function terms in the spectral model. Since each run of the RV pipeline involved 20 iterations, a spread in RV RMS values is generated. Accordingly, we plot each run as a box plot, where for each box the error bars show the full spread in RV RMS values; the vertical extent of the box covers the upper to lower quartile of RV RMS values; and the red line indicates the median RV RMS value. For these tests the blaze function is set as a quadratic polynomial. It is clear that an LSF generated by summing up the first 5 degrees of Hermite functions correspond to the lowest median and absolute RV RMS. This can be understood as the balancing of two effects: for LSFs made up of fewer Hermite functions, the degree of freedom is too low and the LSF is not able to account for all distortions in the line profile. Conversely, LSFs made up of more Hermite functions run the risk of fitting noise in addition to the line profile, diluting the RV information content of the spectral lines. These results are in contrast to previous works that used single Gaussians for their LSFs \citep[]{crockett2011,blake2010,bailey2012,tanner2012}, which corresponds to the 0th degree Hermite function. However, a direct comparison with the latter three works may be inappropriate due to their reliance on a different spectrograph with a reduced spectral resolution. Meanwhile, \citet[]{crockett2011} also used CSHELL, and found no improvements to RV precision by using multi-Gaussian LSFs. It is unknown how their results would change if Hermite function LSFs were used instead. An additional complication is the change in LSF shape with wavelength, which was included in the model of \citet[]{blake2010}. However, this is not an issue for our work due to our short wavelength range (6 nm) compared to theirs (34 nm), such that we can assume the LSF shape is independent of wavelength. 

Figure \ref{fig:gj15a_poly_test} shows the resulting RV precision when using blaze functions constructed with polynomials of order 6 or less in the spectral model, in a similar format as for the LSF tests. The test case with a constant blaze function (``order 0'') is not shown as its RV RMS value is more than 5 times larger than the plotted cases. For these tests the LSF was set as the sum of the first 5 degrees of Hermite functions. Unlike the LSF tests, the best solution is not immediately clear, though the quadratic blaze function provides the lowest minimum and median RV scatter. Blaze functions with order $>$4 increase the RV scatter due to the high sensitivity of the blaze function amplitude to the higher order terms. Previous works have used both linear \citep[]{valenti1995,bailey2012} and quadratic \citep[]{crockett2011} blaze functions, with the choice likely depending on the degree to which the spectra were already normalized to the continuum during the data reduction phase. For example, the California Planet Search RV pipeline uses three order-6 polynomials of decreasing amplitude to normalize their spectra to the continuum before the RV extraction phase, during which only a constant normalization factor is needed (John A. Johnson, 2013, private communications). This method is valid for optical wavelengths where the continuum of the spectrum is well defined between the spectral lines. However, as the spectral lines dominate over the continuum in our wavelength range (Figure \ref{fig:spectra}), higher order blaze functions could have exaggerated amplitudes due to fitting to deep lines or absorption bands, thereby introducing distortions to the spectral model flux when using them for normalization. 

\subsection{RV Stability at High Signal-to-Noise: SV Peg}\label{sec:svpeg_results}

The ensemble RVs of SV Peg and their associated $1\sigma$ error bars are given in Table \ref{table:gj15a_gj876_svpeg_rv} and are shown in Figure \ref{fig:svpeg_allrv}. RV variations with amplitude $\sim$ 1.5 km s$^{-1}$ spanning hundreds of days can be seen, which are consistent with the 145 day variability time scale and amplitude from previous observations \citep[e.g.,][]{hinkle1997} and reaffirms SV Peg's quasi-periodic pulsations as an M giant. Therefore, in order to investigate RV stability using SV Peg observations, we consider its RVs on a shorter time scale. For example, if we assume purely periodic variability, such that the RVs vary by 3 km s$^{-1}$ within half of the total variability period, 145 days, then within 1 hour the RVs will change by only $\sim$ 1.7 $\ms$. The actual change is likely to vary by at least a factor of two due to changes in the RV slope occurring on shorter time scales \citep[see, for example, Figure 8 of][]{hinkle1997}. 

Figure \ref{fig:svpeg_rawrv} shows the individual RVs of SV Peg on five nights when $>$ 100 spectra were taken, each with RV RMS $\sim$ 18\textendash 40 $\ms$. As these values are an order of magnitude larger than the predicted RV variations within a few hours, it is not surprising that no long term trends are apparent in the data. The left column of Figure \ref{fig:svpeg_noisefloor} shows the RV precision as a function of increasing SNR per pixel for each observation, calculated from progressively binning an increasing number of individual RV points, up to one-third of all points obtained that night per bin. The resulting RV noise floors lie at $\sim$2\textendash 6 $\ms$. The SNRs needed to reach the noise floor are $\sim$400\textendash 600 for the 5 nights, which reflects differing weather conditions, pointing/guiding/focusing issues, detector artifacts, and contributions from read noise and dark noise. The theoretical integration time to reach the noise floor for SV Peg, calculated as the product of the integration time per individual exposure ($\sim$3\textendash 5s) and the square of the ratio of the total SNR required ($\sim$400\textendash 600) to the SNR per individual exposure (see Table \ref{table:observations}), is $\sim$200 s. Both the individual and binned RV RMS values are consistent with previous analyses of these data performed by \citet[]{anglada2012}, as well as the level of precision expected given SV Peg's intrinsic RV variability within hours, as calculated in the previous paragraph. In other words, as a result of the high SNR of the observations and the high information content of the deep CO lines of SV Peg stellar spectra, our RV precision for SV Peg may be significantly limited by stellar variability in addition to photon noise or wavelength calibration.

To test this possibility we repeat the calculations shown in the left column of Figure \ref{fig:svpeg_noisefloor} and add an extra step where we subtract a linear trend from the binned RV points at each SNR before calculating $\sigma_{RV}$. The linear trend is fitted to the binned RV points weighted by the inverse square of the scatter in each bin and represents the short-term RV variability due to stellar pulsations. The results are shown in the right column of Figure \ref{fig:svpeg_noisefloor}, where for all but the night of 11/08/20 there is some reduction in the RV slope. However, this is not definitive evidence that we have uncovered the signature of stellar pulsations, as the subtracted linear trend could be due to some other systematic, such as variations in the gravity vector, focus, and/or temperature. Figure \ref{fig:svpeg_linslope} shows the linear RV slope for the subtracted linear trend at each SNR. If the slope is due to pulsation, then they should all have similar magnitudes (a few $\ms$ hr$^{-1}$) and be consistent across many values of the SNR. It is clear that this is not the case for most of the nights, where the magnitudes are either too large (10/10/09, 10/10/10, 10/10/13) and/or too variable (10/10/10, 11/08/20). By contrast, for the night of 11/07/10, the slope's magnitude is consistent with stellar pulsations and is stable across many SNR values. In addition, it features the largest decrease in noise floor upon subtraction of the linear trend, and the total duration of the observations (4 hours) may be sufficient for RV variability due to stellar pulsations to be made apparent. However, more precise RV measurements and complementary observations of activity indicators are needed to fully evaluate whether the linear slope is indeed due to stellar variability.  

\begin{figure}[p]
\centering
\includegraphics[width=0.8\textwidth]{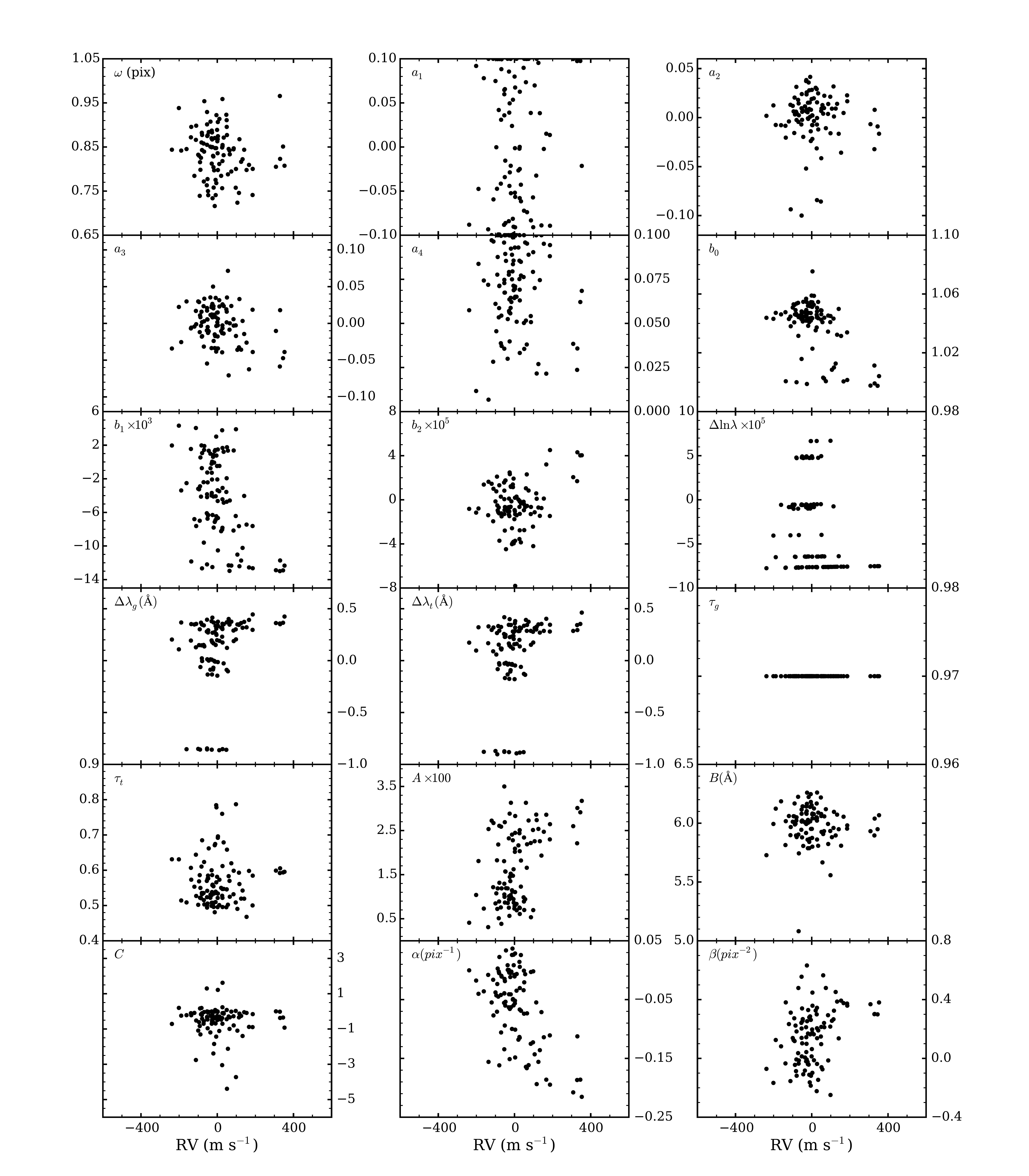}  
\caption{Correlations between the free parameters of the RV extraction pipeline with the individual RV values of GJ 15 A. See Table \ref{table:rvparams} for the definitions of the parameter symbols.}
\label{fig:gj15a_rv_vs_param}
\end{figure}

\begin{figure}[p]
\centering
\includegraphics[width=0.5\textwidth]{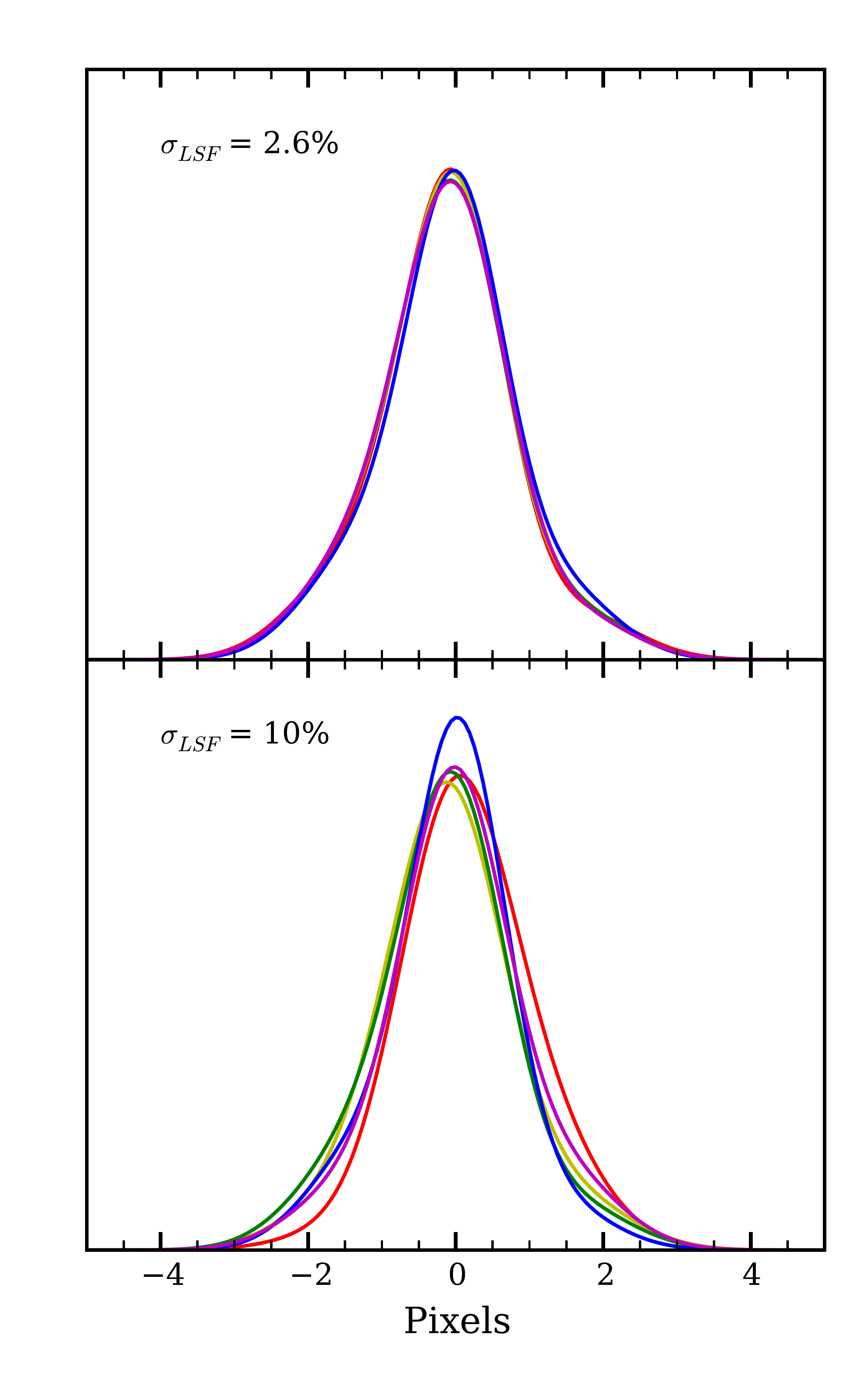}  
\caption{Variability of five GJ 15 A line spread functions within a single night (top) and across several nights (bottom). Different colored lines are used to distinguish between different LSFs. $\sigma_{LSF}$ is defined as the average of the standard deviations of the five LSF magnitudes from a mean LSF within 4 pixels to either side of the zero point on the x axis. The mean LSF is derived from the average of the five LSFs in each case. }
\label{fig:gj15a_lsfvar}
\end{figure}

\begin{figure}[p]
\centering
\includegraphics[width=0.8\textwidth]{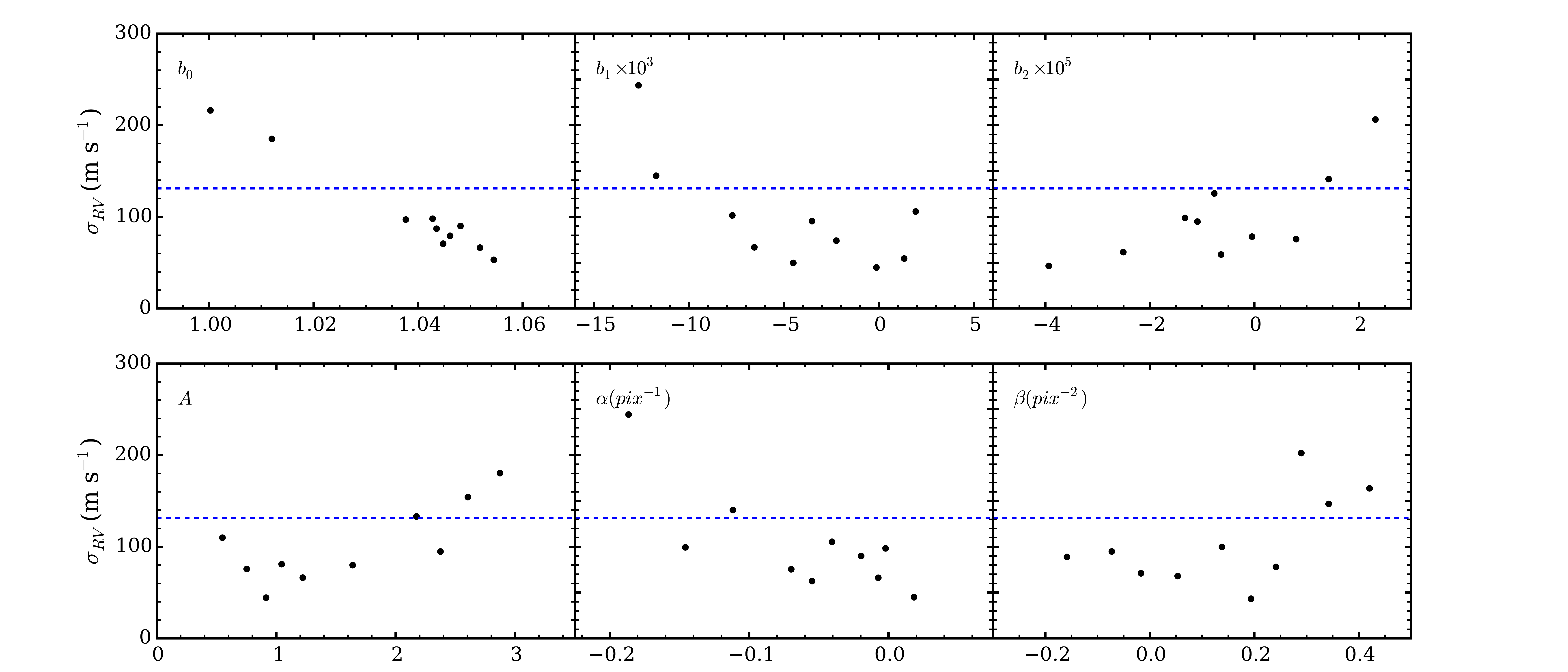}  
\caption{Standard deviations of every 10 individual RV points of GJ 15 A ordered in increasing values of the blaze function coefficients ($b_0$, $b_1$, $b_2$), the fringing amplitude ($A$), and the wavelength solution coefficients ($\alpha$, $\beta$). The standard deviations of all of the individual GJ 15 A RVs are shown by the blue dotted lines.}
\label{fig:gj15a_binned_params}
\end{figure}

\begin{figure}[p]
\centering
\includegraphics[width=0.8\textwidth]{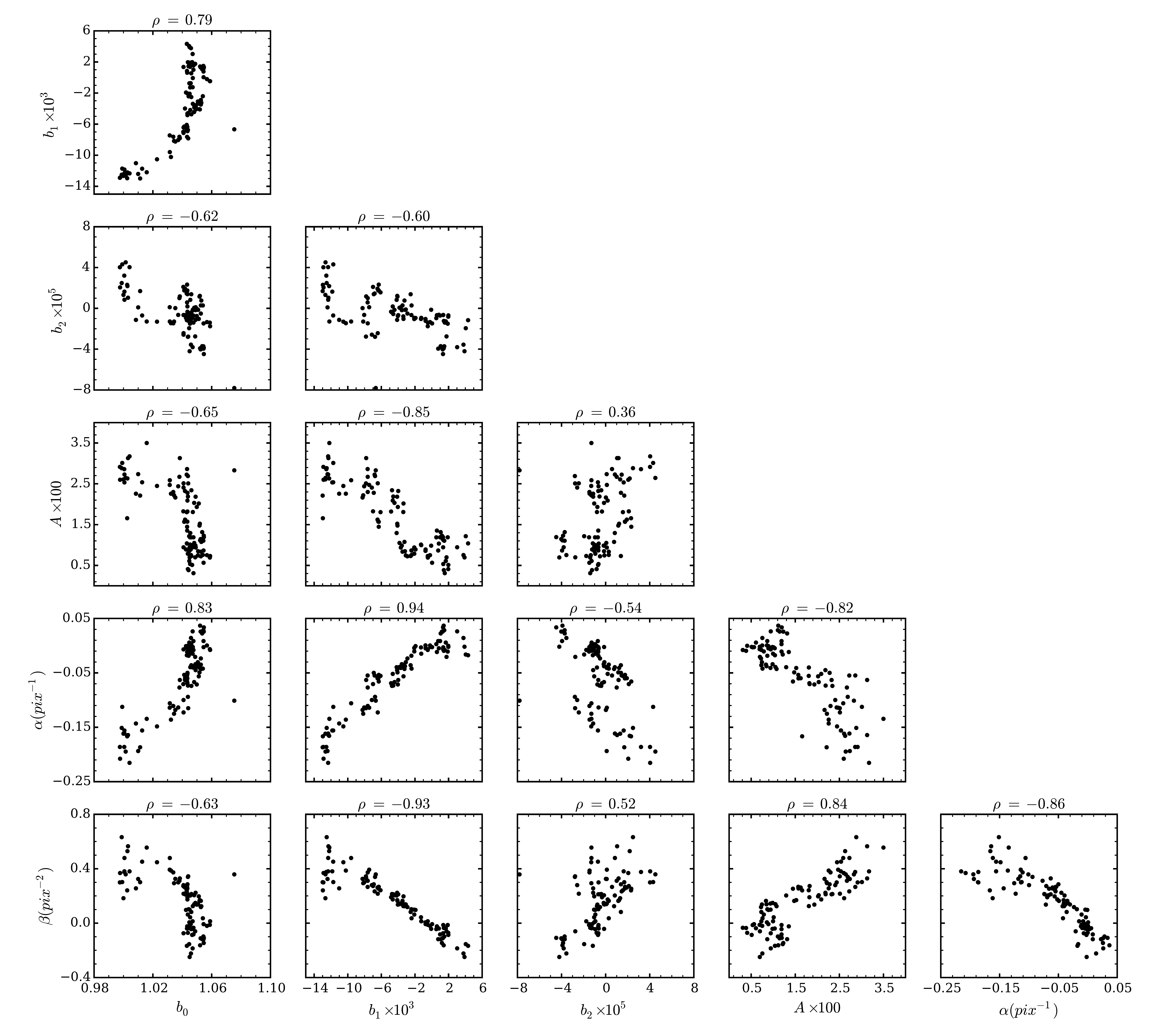}  
\caption{Correlations and Pearson's linear correlation coefficient ($\rho$) between some of the spectral model parameters (constant ($b_0$), linear ($b_1$), and quadratic ($b_2$) terms of the blaze function, the fringing amplitude ($A$), and the linear ($\alpha$) and quadratic ($\beta$) terms of the wavelength solution) from fits to GJ 15 A spectra.}
\label{fig:gj15a_paramcorr}
\end{figure}

\begin{figure}[p]
\centering
\includegraphics[width=0.8\textwidth]{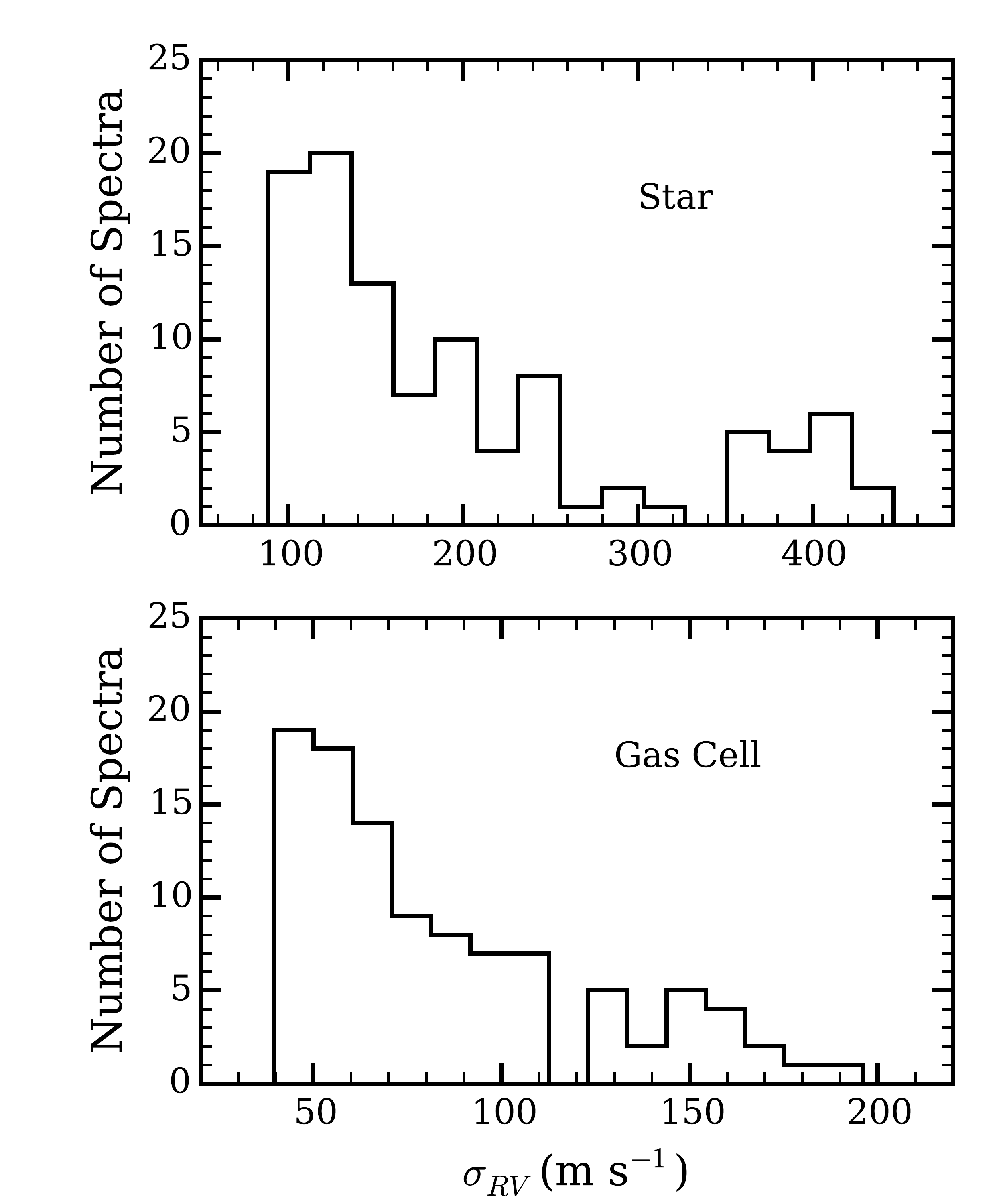}  
\caption{Histograms of expected RV errors in all 103 GJ 15 A spectra from just photon noise (top) and just wavelength calibration (bottom). }
\label{fig:gj15a_rverr}
\end{figure}

\begin{figure}[p]
\centering
\includegraphics[width=0.5\textwidth]{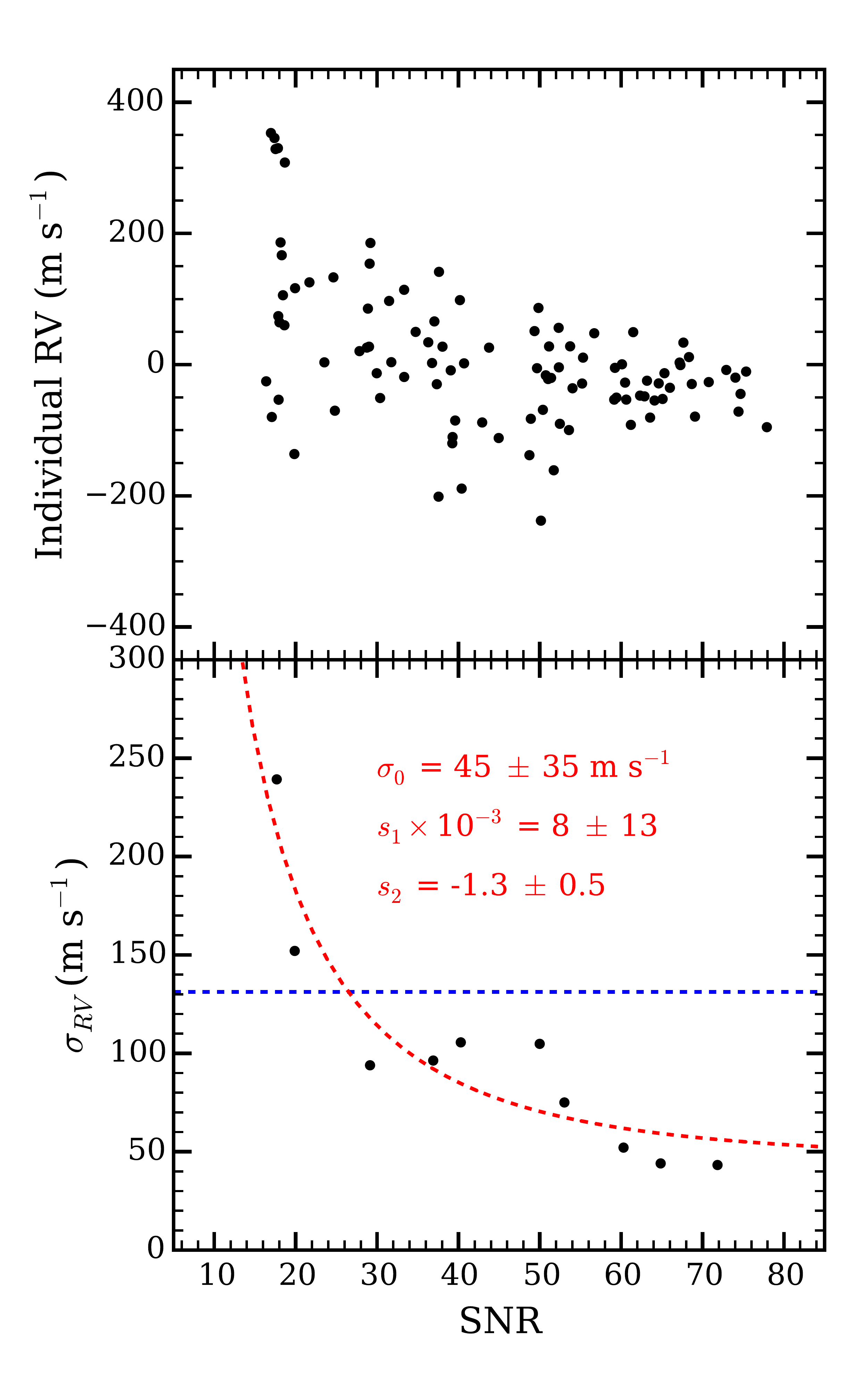}  
\caption{(Top) Distribution of the individual GJ 15 A RVs as a function of the SNR per pixel of each observation. (Bottom) The standard deviation of every 10 individual RV points ordered in increasing values of the SNR. The resulting trend is fitted with a power law (Eq. \ref{eq:sigmarvsnr}, red dotted line), the best fit parameters of which are given in red. The standard deviations of all of the individual GJ 15 A RVs are shown by the blue dotted line.  }
\label{fig:gj15a_rv_vs_snr}
\end{figure}

\begin{figure}[p]
\centering
\includegraphics[width=0.8\textwidth]{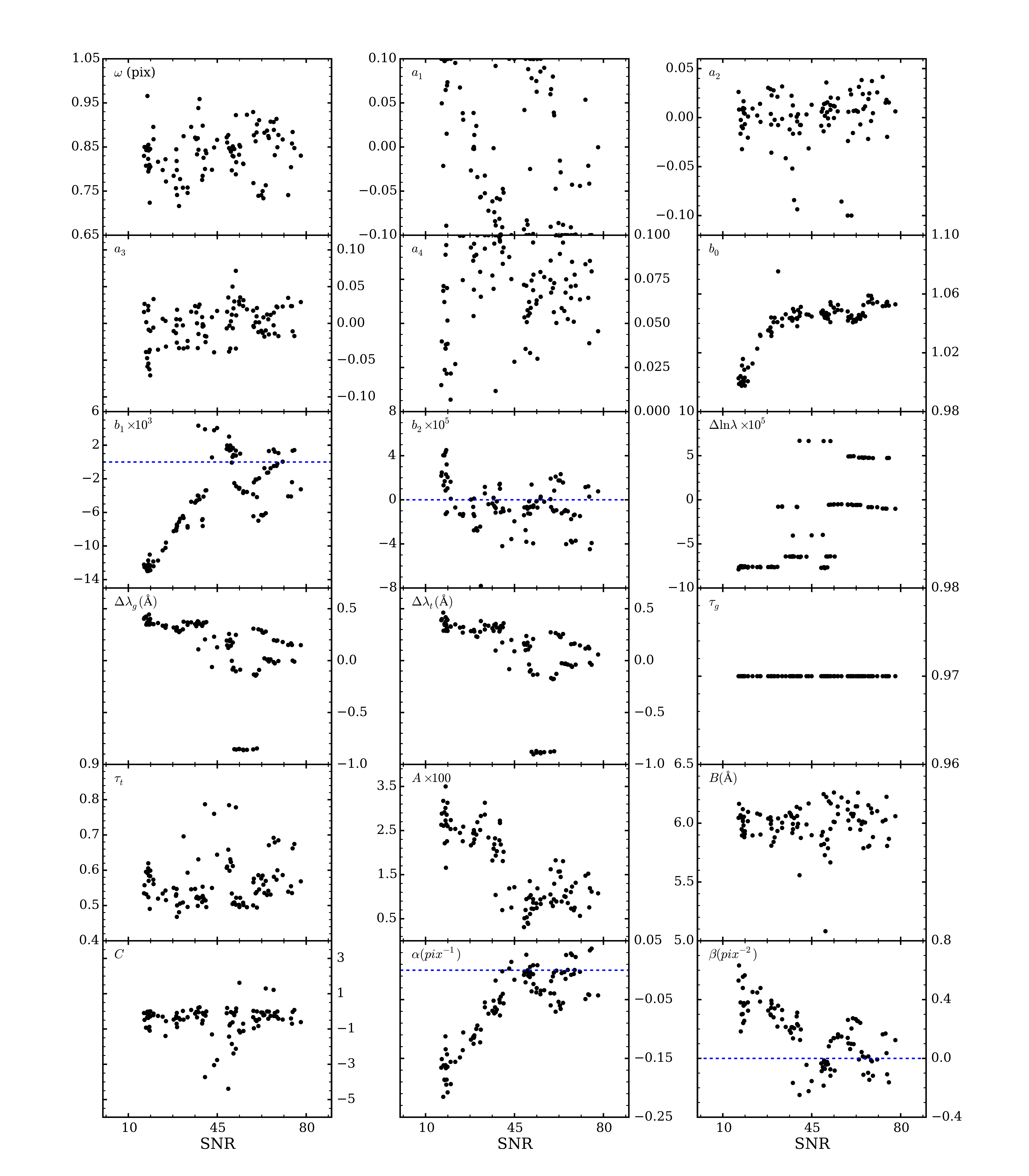}  
\caption{Correlations between the free parameters of the RV extraction pipeline with the SNR per pixel of the individual observations of GJ 15 A. See Table \ref{table:rvparams} for the definitions of the parameter symbols. Blue dotted lines mark the zero values in the $b_1$, $b_2$, $\alpha$, and $\beta$ plots.}
\label{fig:gj15a_params_vs_snr}
\end{figure}

\begin{figure}[p]
\centering
\includegraphics[width=0.5\textwidth]{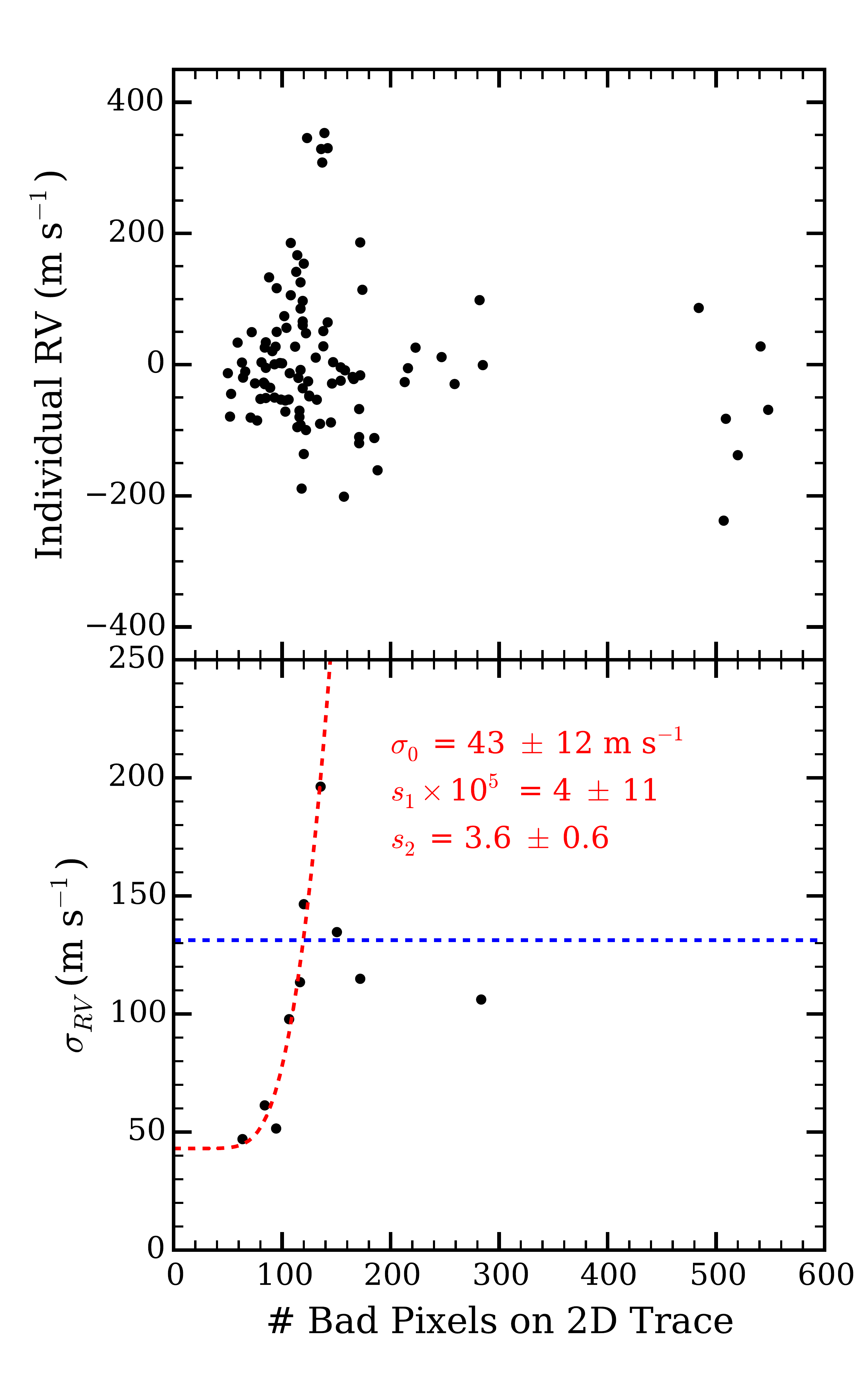}  
\caption{(Top) Correlation between the individual GJ 15 A RVs and the number of bad pixels on the 2D spectral trace of each observation. (Bottom) The standard deviation of every 10 individual RV points ordered in increasing numbers of bad pixels. The resulting trend is fitted with a power law (Eq. \ref{eq:sigmarvsnr}, with N$_{bp}$ in place of the SNR, shown by the red dotted line), the best fit parameters of which are given in red. The standard deviation of all of the individual GJ 15 A RVs are shown by the blue dotted lines.}
\label{fig:gj15a_rv_vs_bp}
\end{figure}

\begin{figure}[p]
\centering
\includegraphics[width=0.8\textwidth]{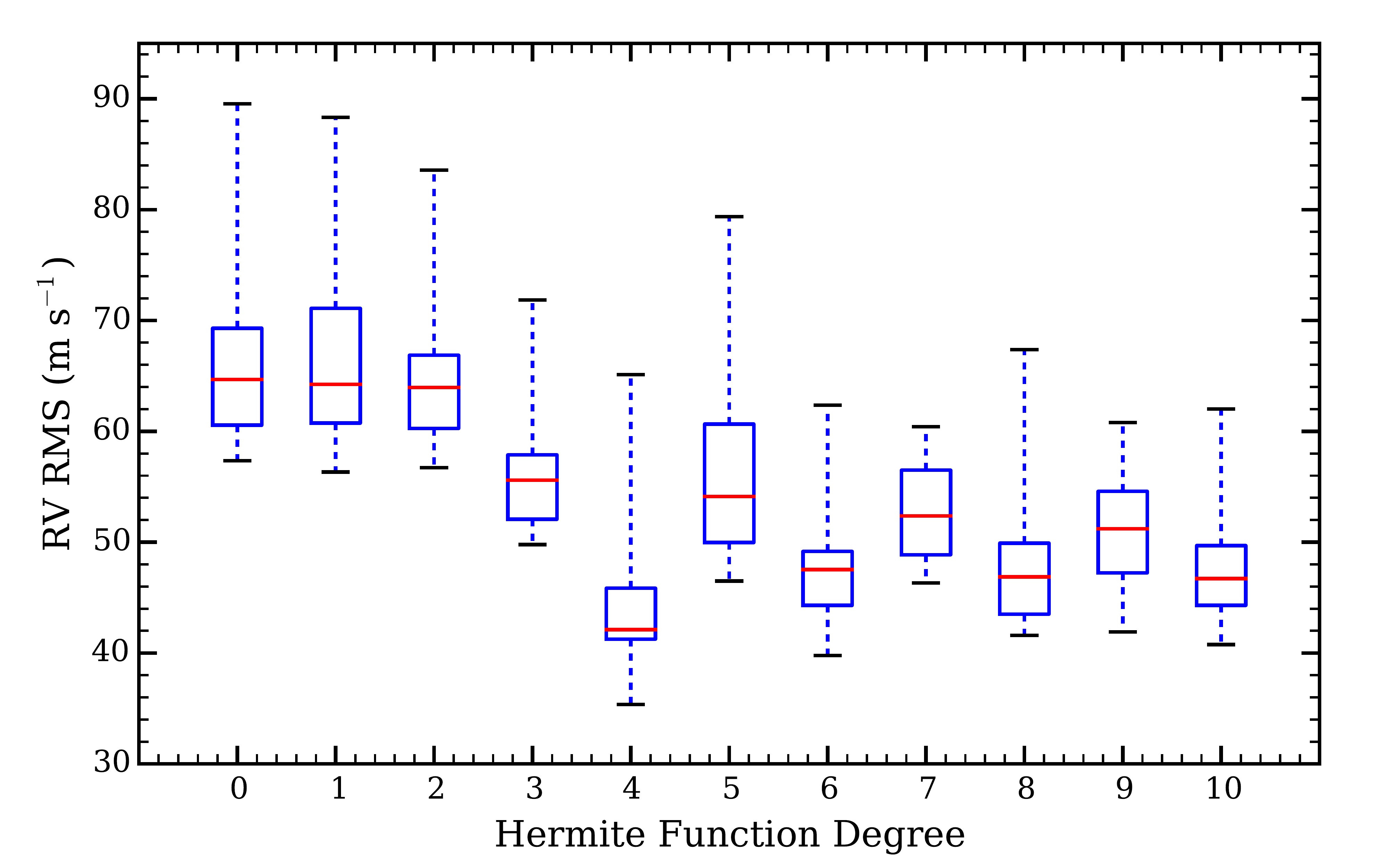}  
\caption{Sensitivity of the GJ 15 A RV RMS to the degrees of Hermite functions used to construct the spectral model LSF. Each test case was run for 20 iterations, with each box plot representing the full range of RV RMSs for all 20 iterations of each case. The upper and lower bounds of the error bars show the highest and lowest RV RMSs achieved during each test case, respectively; the upper and lower bounds of the blue boxes show the first and third quartile RV RMSs for each set of 20 iterations for each case, respectively; and the red horizontal line in each blue box show the median RV RMS of each case.}
\label{fig:gj15a_lsf_test}
\end{figure}

\begin{figure}[p]
\centering
\includegraphics[width=0.8\textwidth]{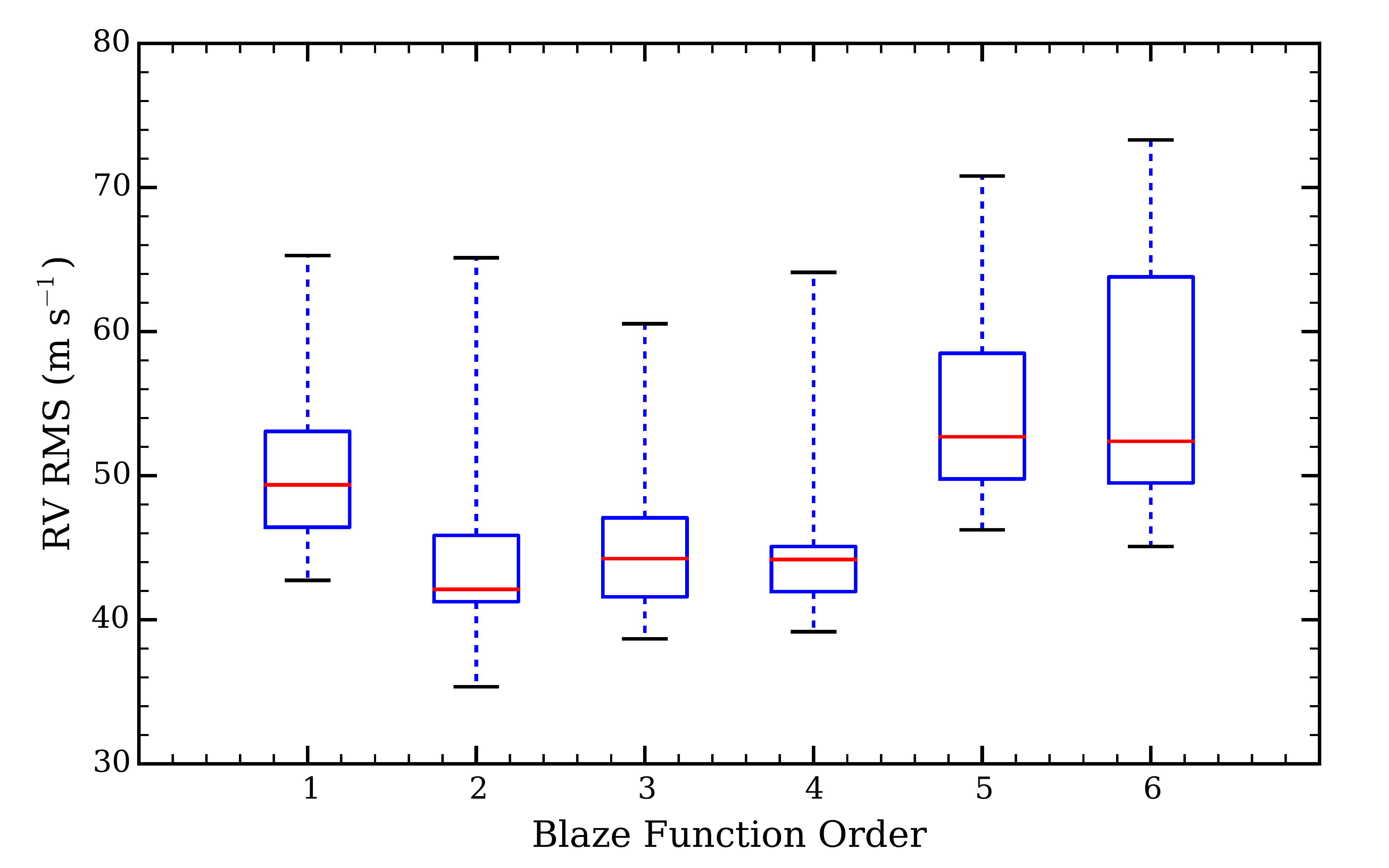}  
\caption{Same as Figure \ref{fig:gj15a_lsf_test}, but for the sensitivity of the GJ 15 A RV RMS to the orders of polynomials used to construct the spectral model blaze function.}
\label{fig:gj15a_poly_test}
\end{figure}

\begin{figure}[p]
\centering
\includegraphics[width=0.8\textwidth]{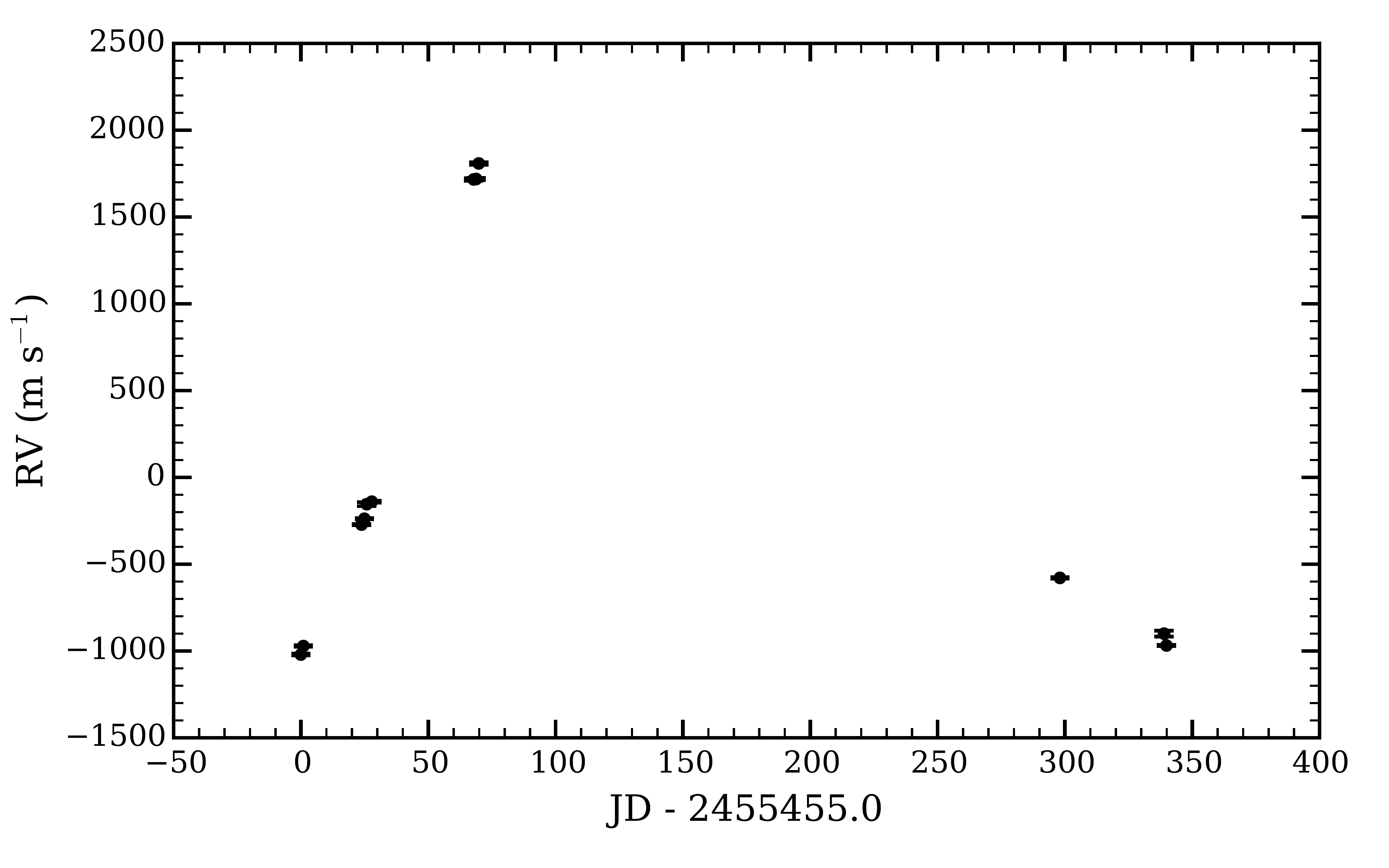}  
\caption{The nightly RVs of SV Peg. }
\label{fig:svpeg_allrv}
\end{figure}

\begin{figure}[p]
\centering
\includegraphics[width=0.5\textwidth]{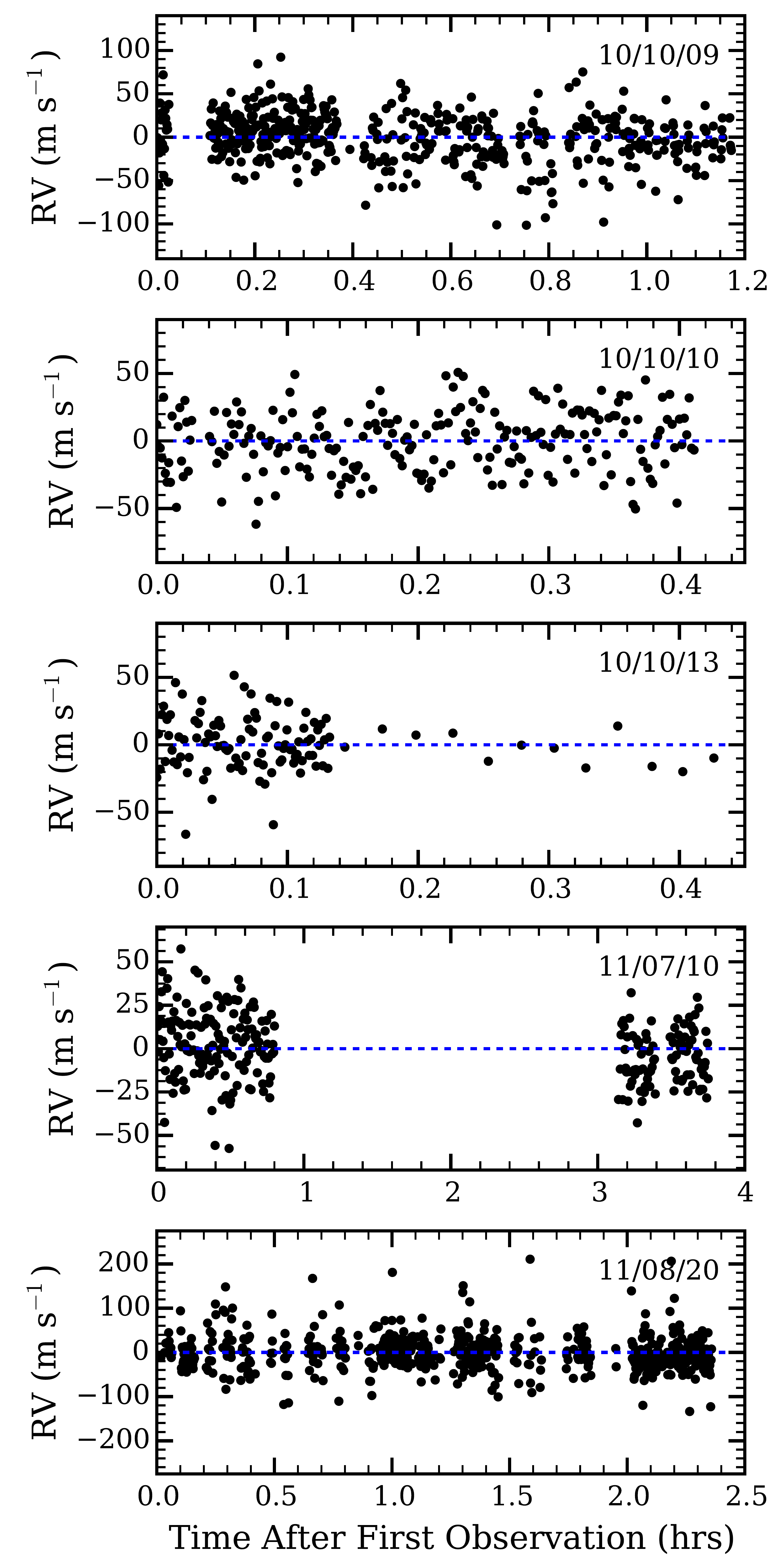}  
\caption{Individual SV Peg RVs on the indicated nights. The date format is year/month/day.  }
\label{fig:svpeg_rawrv}
\end{figure}

\begin{figure}[p]
\centering
\includegraphics[width=0.8\textwidth]{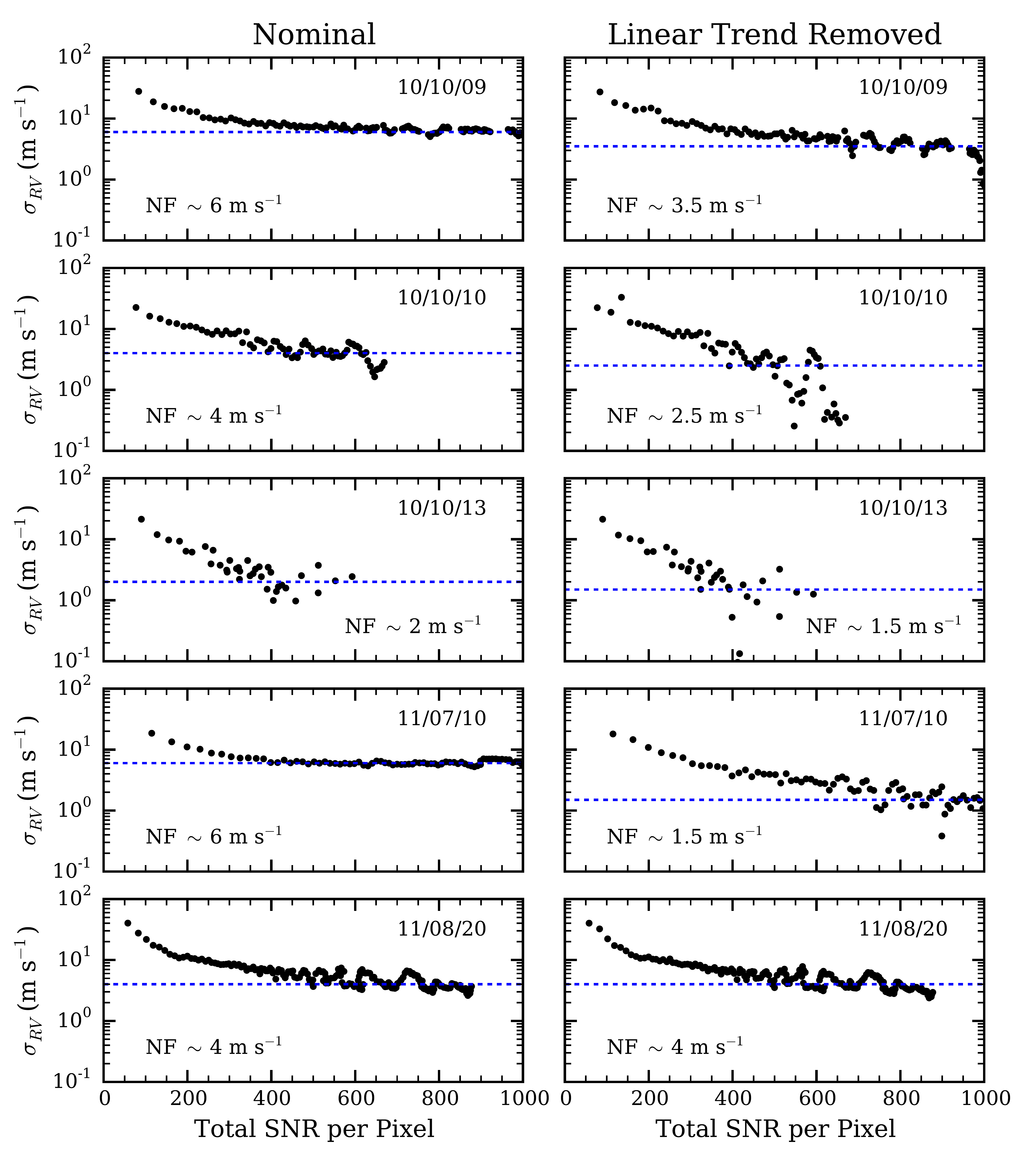}  
\caption{(Left) RV noise floors (NF) of the indicated nights, calculated by binning progressively more individual RV points and calculating the standard deviation of the resulting binned RV points. The blue dotted lines mark the approximate RV noise floors, as indicated by the asymptotic value of $\sigma_{RV}$ of the binned RV points. (Right) Same as the left column, but $\sigma_{RV}$ is calculated after a linear trend is subtracted from the binned RVs.}
\label{fig:svpeg_noisefloor}
\end{figure}

\begin{figure}[p]
\centering
\includegraphics[width=0.5\textwidth]{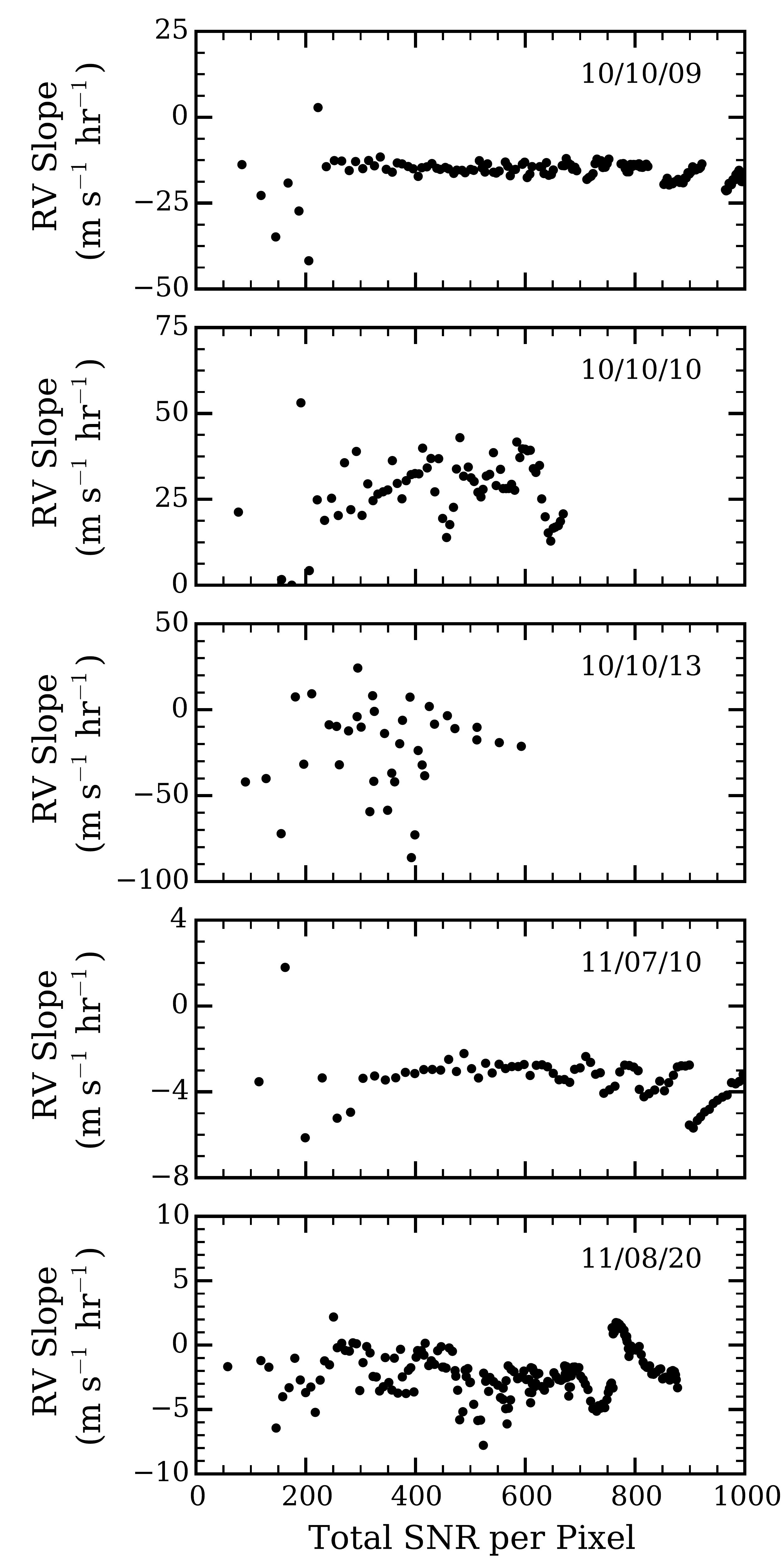}  
\caption{The slope of the linear trend subtracted from the binned RVs used to calculate the noise floors shown in Figure \ref{fig:svpeg_noisefloor}, for the indicated nights.}
\label{fig:svpeg_linslope}
\end{figure}

\begin{figure}[p]
\centering
\includegraphics[width=0.8\textwidth]{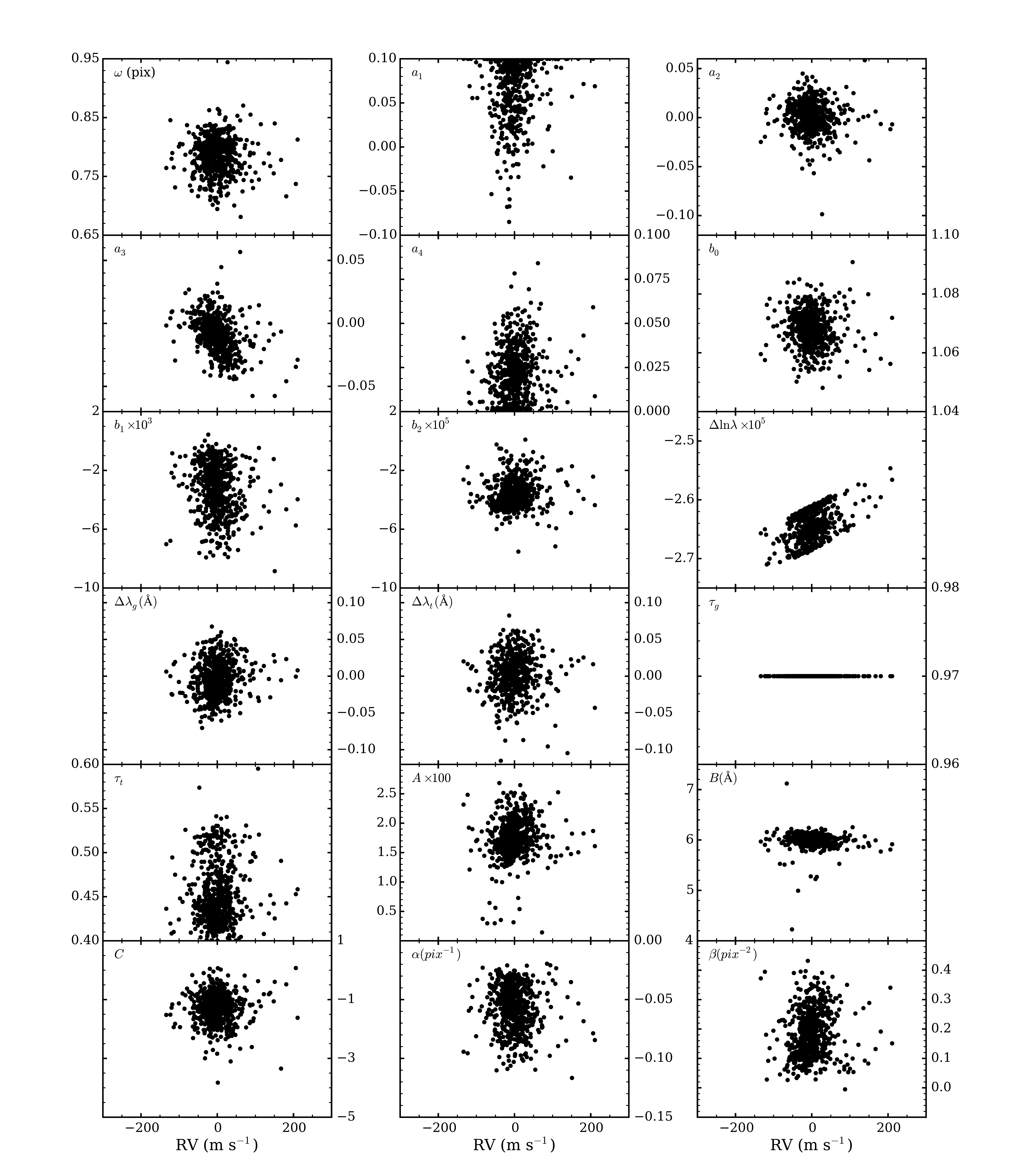}  
\caption{Same as Figure \ref{fig:gj15a_rv_vs_param}, but for SV Peg observations taken on 11/08/20. Note the different scales on the axes.}
\label{fig:svpeg110820_rv_vs_param}
\end{figure}

Figure \ref{fig:svpeg110820_rv_vs_param} shows the correlation between RV and the RV pipeline parameters for SV Peg observations taken on August 20th, 2011. This date was chosen as it had the most RV points. These results show much less structure than those shown in Figure \ref{fig:gj15a_rv_vs_param}, which is due to the much shorter time baseline of the observations. Like for GJ 15 A, no obvious correlations exist, except for the relative wavelength shift, $\Delta{\ln{\lambda}}$. This is due to scatter in the retrieved RVs of SV Peg. To illustrate, suppose the retrieved RVs are made up of two components: a scattered component, and an ideal component that, when added to the barycentric correction, results in zero RVs. Therefore, the barycenter-corrected RVs would just be made up of this scattered component, and thus they are linearly correlated with each other. The same correlation exists in the analogous panel of Figure \ref{fig:gj15a_rv_vs_param}, but is isolated to single epochs.

\begin{figure}[p]
\centering
\includegraphics[width=0.8\textwidth]{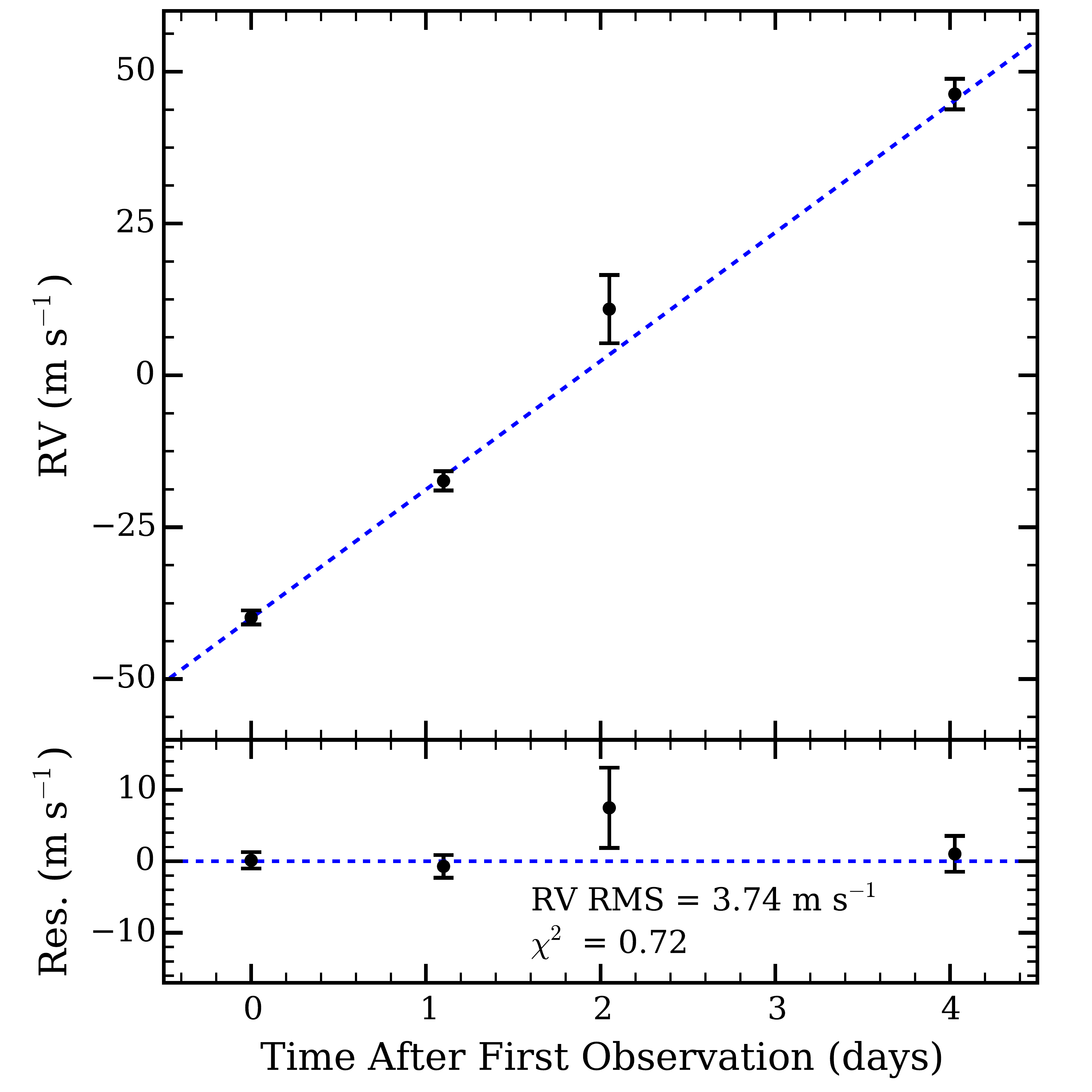}  
\caption{(Top) SV Peg nightly RVs from the October 2010 run, with the best-fit linear trend overplotted (blue dotted line). (Bottom) The residual RVs after subtraction of the linear trend.}
\label{fig:svpeg1010_rv}
\end{figure}

\begin{figure}[p]
\centering
\includegraphics[width=0.8\textwidth]{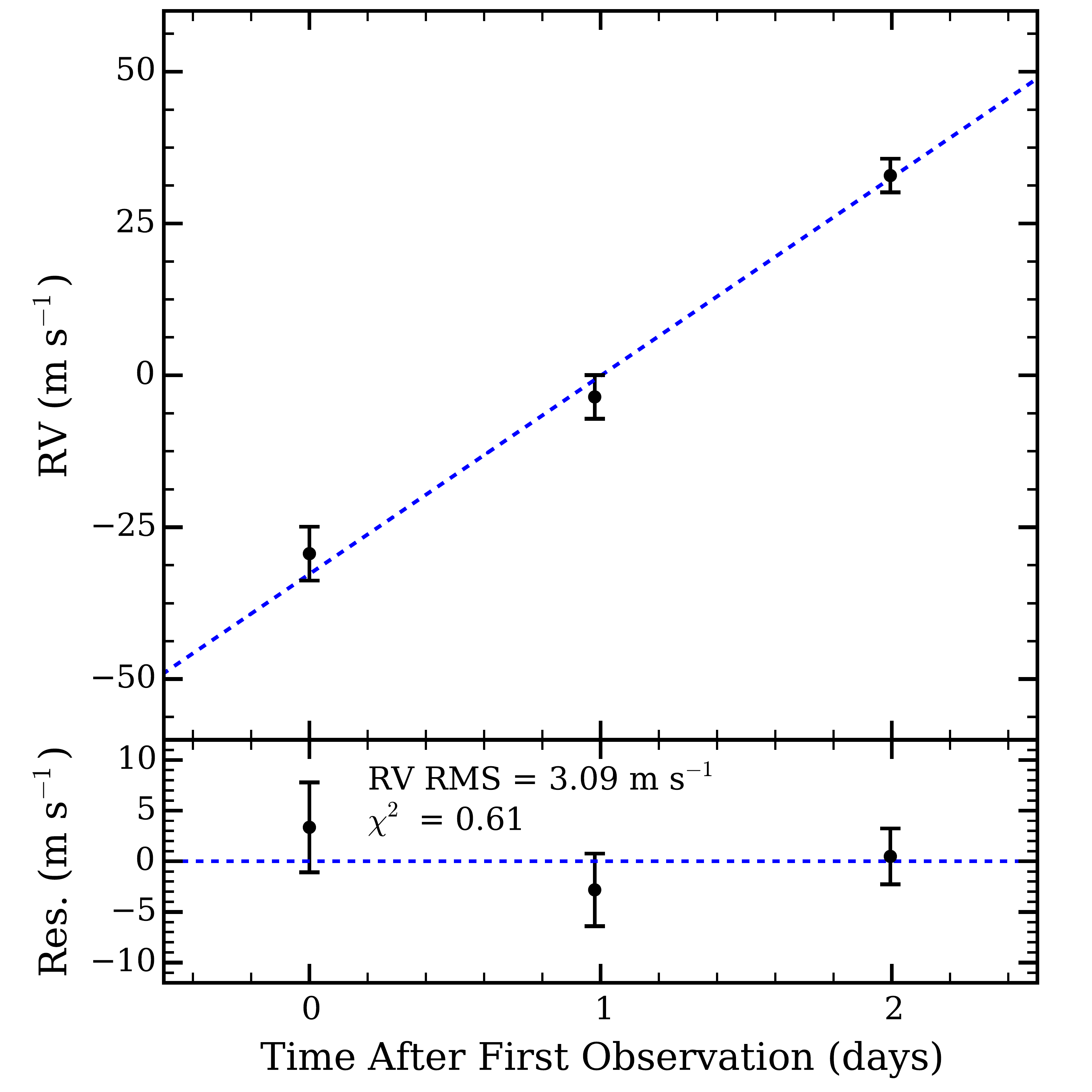}  
\caption{Same as Figure \ref{fig:svpeg1010_rv}, but for the November 2010 run. }
\label{fig:svpeg1011_rv}
\end{figure}

Figures \ref{fig:svpeg1010_rv} and \ref{fig:svpeg1011_rv} show the nightly RVs for two separate observing runs, one in October 2010 and one in November 2010, respectively. Each run consists of spectra taken on nearly consecutive nights, revealing a short-term linear RV trend resulting from SV Peg's intrinsic variability. Given the $\sim$ 1.7 $\ms$ per hour RV slope calculated above, over several days we expect to observe RV changes of $\sim$ 100 $\ms$, roughly consistent with the slopes shown in Figures \ref{fig:svpeg1010_rv} and \ref{fig:svpeg1011_rv}. Once the linear trends are subtracted, the resulting RV scatter is $\sim$ 3\textendash 4 $\ms$, similar to the noise floors of the intra-night results. The detrended nightly RVs of the Oct. 2010 and Nov. 2010 runs and their associated $1\sigma$ error bars are given in Table \ref{table:svpeg_rv}. Both our intra-night and intra-run results are consistent with the observed RV stability of other M giant stars on similar time scales \citep[e.g.,][]{seifahrt2008}. 

\begin{figure}[p]
\centering
\includegraphics[width=0.8\textwidth]{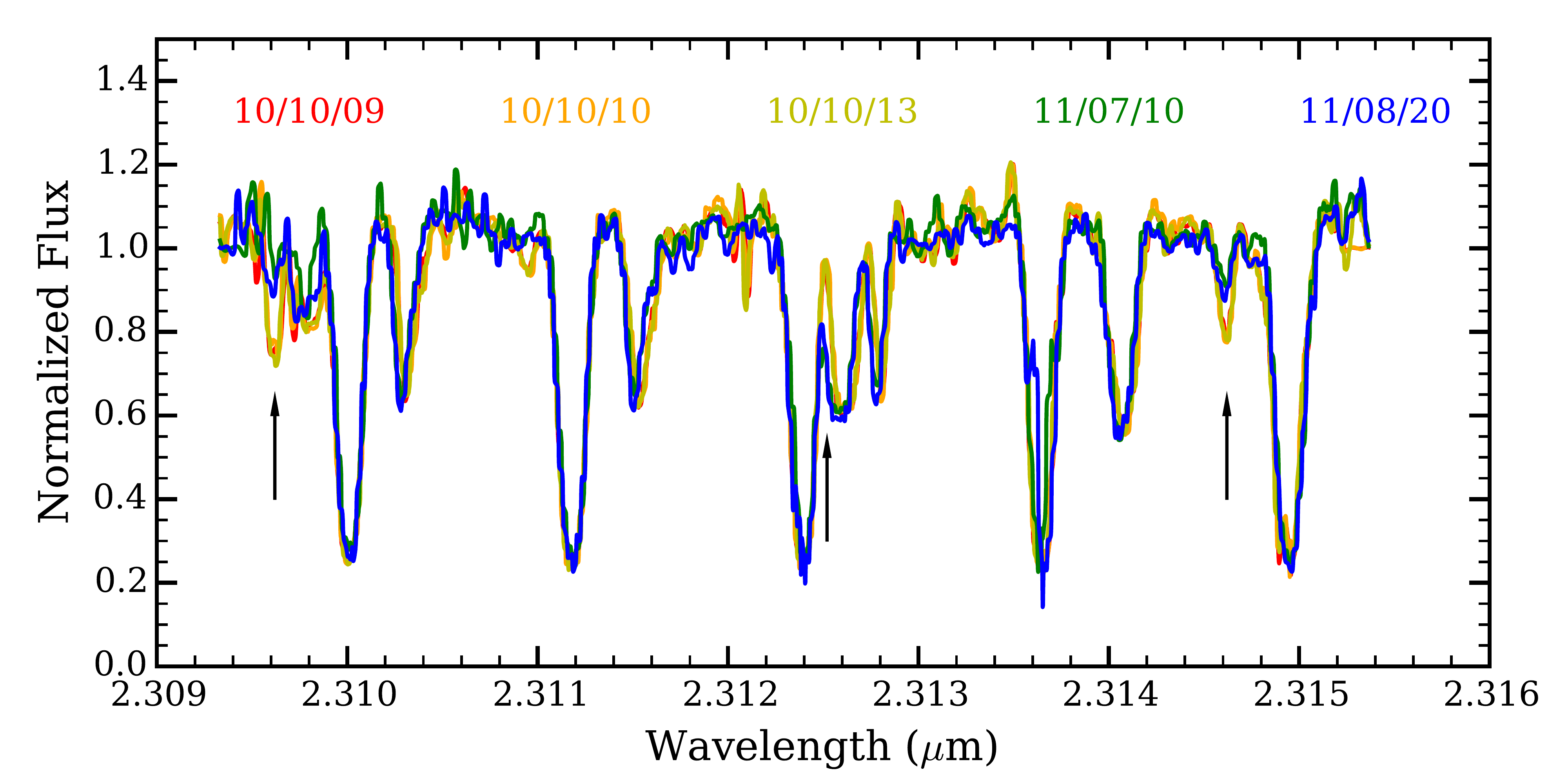}  
\caption{Comparison of separately retrieved stellar templates from the five indicated nights. Major deviations between the templates are indicated by the arrows. }
\label{fig:svpeg_templates}
\end{figure}

The error bars for the SV Peg nightly RVs shown in Table \ref{table:svpeg_rv} are smaller than those of the same nightly RVs shown in Table \ref{table:gj15a_gj876_svpeg_rv}. This results from the different sets of spectra input into the RV extraction pipeline for these two cases, and the way we derive our stellar templates. In Table \ref{table:gj15a_gj876_svpeg_rv}, all SV Peg spectra were input simultaneously, deriving one single stellar spectrum used to fit all observed spectra despite SV Peg's variability; in Table \ref{table:svpeg_rv}, only the spectra from the individual runs were input, and separately, thus producing two different stellar spectra for the two runs. The smaller error bars for the nightly RVs in Table \ref{table:svpeg_rv} then suggest that the nightly RV scatter is smaller when using a stellar template derived specifically from the spectra within those runs, rather than one derived using all available SV Peg spectra. In other words, the stellar spectrum of SV Peg is changing over its long-term variability cycle such that no one stellar template can be used to satisfactorily fit all observations. We explore this effect in Figure \ref{fig:svpeg_templates}, where five stellar templates derived from the five separately analyzed nights shown in Figures \ref{fig:svpeg_rawrv} and \ref{fig:svpeg_noisefloor} are compared. Some small differences are expected due to noise in the observations, but larger differences are also present, as marked by the arrows. The wavelengths of these large differences do not correspond to any major gas cell or telluric features and thus are likely due to changes in SV Peg itself. We also do not see this phenomenon in any of the other targets analyzed using our RV extraction pipeline \citep[]{gagne2015}. Furthermore, the two groups of templates, one derived from 2010 data and one derived from 2011 data, show greater differences between them than between templates within each group, indicating, as expected, that the stellar spectrum changed more over months than over days. Although a full interpretation of these results is beyond the scope of this paper, it would be interesting to ascertain what these changes to the stellar spectrum mean for the chemistry and dynamics of the atmosphere of SV Peg. 

\subsection{Validation of Planet-Detection Capabilities: GJ 876}\label{sec:gj876_results}

\begin{figure}[p]
\centering
\includegraphics[width=0.8\textwidth]{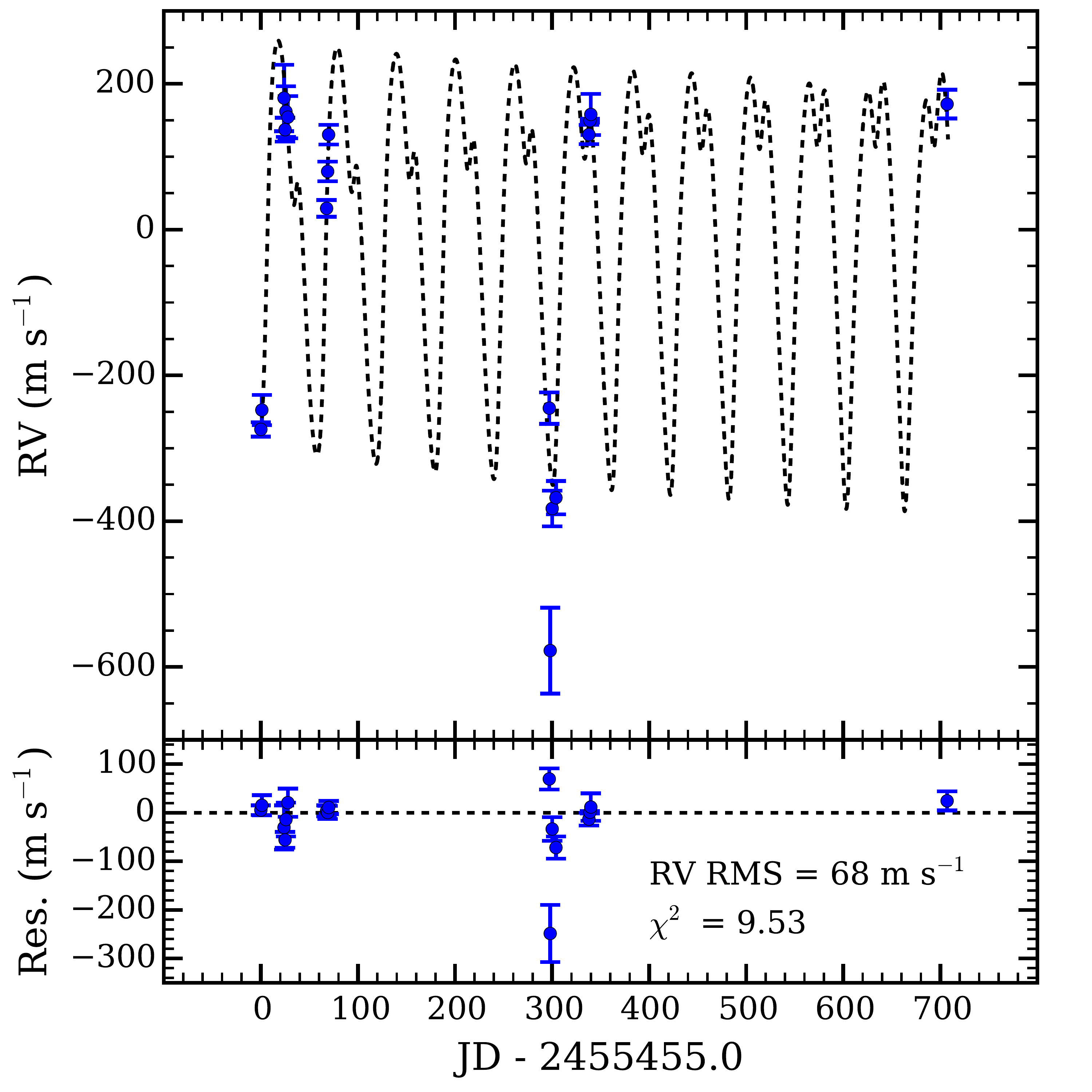}  
\caption{(Top) The nightly RVs of GJ 876 (blue points) and a best-fit 2-planet RV solution curve optimized using Systemic Console (black dotted line). (Bottom) Residual RVs after subtraction of the RV solution curve from the nightly RVs. Removal of the outlier at JD\textendash 2455455 $\sim$ 300 reduces the RMS of the residuals to 35 $\ms$.}
\label{fig:gj876_rv}
\end{figure}

The top panel of Figure \ref{fig:gj876_rv} shows our GJ 876 nightly RVs (blue), while Table \ref{table:gj15a_gj876_svpeg_rv} gives their values and their associated $1\sigma$ error bars. The amplitude of the RV variations are consistent with those observed in optical surveys, aside from the outlier at JD\textendash 2455455.0 $\sim$ 300 with the large error bars \citep[]{delfosse1998,marcy1998,marcy2001,rivera2005,rivera2010}. Using the Systemic Console \citep[]{meschiari2009}, \citet[]{gagne2015} was able to fit a 1-planet solution to the data without any prior constraints and retrieve orbital parameters within 2$\sigma$ of the published values for planet b, thereby confirming, for the first time, the existence of GJ 876 b using NIR RVs. 

To further test the pipeline results, we fit a 2-planet solution to our data using the Systemic Console, corresponding to planets b and c. The best fit RV curve is shown in the top panel of Figure \ref{fig:gj876_rv} (dashed line). Due to the low cadence of our data and the rapid dynamical evolution of the system, it is difficult to accurately retrieve the orbital parameters of both of these planets. Instead, we fix the planet masses, orbital periods, and orbital eccentricities to the published values and allow the mean anomaly, longitude of periastron, and the velocity zero point to vary. Our best fit solution does not show the alignment of the two planets' longitudes of periastron reported in previous studies \citep[e.g.,][]{laughlin2005,rivera2005}, but that is likely due to the low cadence of our observations. The fit results in a residual RMS of 68 $\ms$ (bottom panel of Figure \ref{fig:gj876_rv}); removing the outlier at JD\textendash 2455455.0 $\sim$ 300 reduces this value to 35 $\ms$, within the RV precision of our RV standard, GJ 15 A. We do not pursue 3-planet and 4-planet solutions, as the two lower mass planets in the system are beyond the detection limits of our pipeline. 

These results demonstrate that our RV pipeline is capable of detecting Jupiter-mass planets around M dwarfs with orbital periods of tens of days using CSHELL. However, it is clear that a confirmed detection will require high cadence, dedicated observations of RV variables in order to better constrain the orbital parameters of any possible planets they may host. 

\section{Summary and Outlook}\label{sec:discussion}

We have constructed a data analysis pipeline that can process into RVs the NIR spectra of M dwarfs observed using CSHELL on NASA IRTF with the aid of a methane isotopologue gas cell by optimizing fits between the observed spectra and a spectral model. The pipeline is able to retrieve the stellar spectrum in an iterative process directly from the science observations without the need for additional observations, and takes into account temporal variations in the instrumental LSF, curvature in the continuum (blaze function), sinusoidal fringing, and wavelength solution. The pipeline is capable of (1) obtaining an RV precision $\sim$ 35 $\ms$ for the RV standard M dwarf GJ 15 A, with the error contributions coming mostly from photon noise,  (2) obtaining, for the high SNR target, SV Peg, noise floors of $\sim$ 2\textendash 6 $\ms$ on individual nights, and $\sim$ 3 $\ms$ RV precision over several days after subtracting out a linear trend consistent with RV variability due to stellar pulsations, and (3) detecting/confirming at least one Jupiter mass planet around the M dwarf GJ 876 in the NIR. These results are derived from observations with a 22-year old InSb detector within CSHELL, a spectrograph that was not designed with precision NIR RVs in mind, and which is mounted at the Cassegrain focus, thereby introducing significant mechanical, thermal, and pressure variations that all contribute to the RV scatter. Furthermore, the RV information content is restricted to five CO lines across a single order of 256 pixels that covers only 6 nm \citep[]{greene1993}. However, despite these limitations, our results compare favorably to previous works that use larger telescopes, newer detectors, and spectrographs with greater mechanical, thermal, and pressure stability. Therefore, it will be useful to evaluate the possible improvements to our RV stability resulting from the reduction or elimination of the aforementioned limitations in newer NIR spectrographs. We will base our discussions on iSHELL, as it is due to replace CSHELL in the near future. 

iSHELL is a cross-dispersed high resolution echelle spectrograph with R $\sim$ 70000 at minimum slit width (0\farcs375) \citep[]{rayner2012}. As with CSHELL, it will be fixed to the Cassegrain focus of the 3 m telescope at NASA IRTF. It uses a Teledyne 2048x2048 Hawaii 2RG array as its main detector and enables multiple orders to be observed simultaneously. A single exposure in the K band allows for $\sim$ 200 nm to be captured, thereby improving our wavelength coverage by a factor of 200/6 = 33; extrapolating from our calculations in ${\S}$\ref{sec:gj15a_err} with the assumption that the RV information content stays roughly constant with wavelength, this results in a typical $\sigma_{RV}$ = 130/$\sqrt{33}$ $\sim$ 22.5 $\ms$ for the individual RVs due to photon noise and wavelength calibration errors. Given an SNR per pixel of 50 for each observation, and a total SNR of 200 per night, the nightly RV scatter would thus be 22.5/(200/50) $\sim$ 5.6 $\ms$. This is consistent with the estimates of \citet[]{reiners2010} and \citet[]{bottom2013} and will facilitate the detection of Super-Earths and Mini-Neptunes in the Habitable Zones of M dwarfs. 

On the other hand, while the higher spectral resolution of iSHELL will increase the RV information content of the observed spectra, the SNR per pixel will decrease at the same time due to spreading out the same number of photons across more pixels. The impact of this effect on the RV precision depends on the sharpness of the spectral features of the target, which is a function of its rotational velocity. A target that rotates quickly ($vsini$ $\sim$ 10 km s$^{-1}$) may be insensitive to a change of R from 46000 to 70000, while a target that rotates slowly ($vsini$ $<$ 1 km s$^{-1}$) may see its RV scatter increase by 50$\%$ \citep[]{bouchy2001}. In addition, the increase in wavelength coverage will cause severe contamination of the observed spectra by atmospheric water and methane telluric absorption lines and OH emission lines, which are outside of the 6 nm wavelength range of our current CSHELL observations. \citet[]{bean2010} corrected for telluric contamination by modeling the absorption and emission lines using line lists from the HITRAN database in conjunction with a time-resolved model of the atmosphere above the observatory and the LBLRTM radiative transfer code. Further improvements to the modeled telluric lines were made by fitting to a telluric standard star in order to determine systematic wavelength shifts and strength variations of the lines compared to their default values in the HITRAN database. A similar method can be applied to iSHELL observations, with additional improvements to the synthetic telluric template made possible by iterative processes akin to what we have done to improve the gas cell template (${\S}$\ref{sec:spectralmodel}), though it will be more challenging in this case due to the presence of multiple components (methane and water absorption lines and OH emission lines). Alternatively, a synthetic telluric spectrum can be constructed from principle component analysis using a library of observations of telluric standard stars, which was shown by \citet[]{artigau2014} to improve RV precision in the R band without the addition of spurious RV signals, though it remains to be seen how well this method works in the H and K bands. Finally, the deepest telluric lines can simply be masked during RV extraction, though this reduces the RV information content of the observations. 


There exist several factors in addition to the increased wavelength coverage, however, that will help increase the RV precision of targets observed using iSHELL. The upgrades to the filter selection wheel will largely eliminate the sinusoidal fringing and thus its contributions to the RV error. The newer detector will lead to a decrease in the number of bad pixels on the 2D trace, further improving the RV precision by increasing the available information content and reducing the need to interpolate along the spatial direction during data reduction. The increased efficiency of the detector will allow for higher SNR for the same targets currently observable using CSHELL, while also making it possible to observe dimmer targets. Recent tests of iSHELL showed that an SNR per pixel of 100 for targets with K magnitude $\sim$ 9.5 is achievable within one hour (Peter P. Plavchan, 2015, private communications). Comparisons with the targets of \citet[]{gagne2015} show that this will permit future RV surveys with NASA IRTF to observe targets 2\textendash 3 K magnitudes dimmer, thereby increasing the sample size of potential RV targets. Additional upgrades to iSHELL, such as the implementation of fiber feeds for improved stabilization of detector illumination and a laser comb for higher-fidelity wavelength calibration could be possible, and have already been demonstrated using CSHELL \citep[]{plavchan2013b,yi2015}. Given the proliferation of precision NIR RV surveys in the near future, a survey undertaken using iSHELL will nicely complement these efforts. 

\bigskip
 
We thank K. Sung, S. Crawford, B. Drouin, E. Garcia-Berrios, N. S. Lewis, S. Mills, and S. Lin for their effort in the building and setting up of the methane isotopologue gas cell. We thank B. Walp for his help with data collection at NASA IRTF. We thank J. Rayner, L. Bergknut, B. Bus, and the telescope operators at NASA IRTF for their help throughout this project. This work uses observations obtained at NASA IRTF through programs number 2010B022, 2011A083, 2011B083, and 2012B021. This work was supported in part by a JPL Research and Technology Development Grant and the JPL Center for Exoplanet Science. Additional support includes the Venus Express program via NASA NNX10AP80G grant to the California Institute of Technology, and an NAI Virtual Planetary Laboratory grant from the University of Washington to the Jet Propulsion Laboratory and California Institute of Technology under solicitation NNH12ZDA002C and Cooperative Agreement Number NNA13AA93A. The authors recognize and acknowledge the very significant cultural role and reverence that the summit of Mauna Kea has always had within the indigenous Hawaiian community. We are most fortunate to have the opportunity to conduct observations from this mountain.



\clearpage

\clearpage

\clearpage

\clearpage

\clearpage

\clearpage

\clearpage

\clearpage

\clearpage

\clearpage

\clearpage

\clearpage

\clearpage

\clearpage

\clearpage

\clearpage

\clearpage

\clearpage

\clearpage

\clearpage

\clearpage

\clearpage

\clearpage

\clearpage

\clearpage

\clearpage

\clearpage

\clearpage


\begin{deluxetable}{lcccccc}
\tabletypesize{\scriptsize}
\tablewidth{0pt}
\tablecolumns{7}
\tablecaption{Properties of Target Stars Used for RV Pipeline Testing}
\tablehead{
\colhead{Star} & \colhead{R.A.} & \colhead{Decl.} & \colhead{Spectral Type} & \colhead{K mag.} & Reference & \colhead{Notes}}
\startdata
GJ 15 A	&	00 18 22.9	&	+44 01 22.6	&	M2.0V	&	4.02		& \citet[]{jenkins2009}	&	RV Standard	\\
SV Peg	&	22 05 42.1	&	+35 20 54.5	&	M7		&	-0.55		& \citet[]{ducati2002}		&	High SNR Target	\\
GJ 876	&	22 53 16.7	&	-14 15 49.3	&	M5.0V	&	5.01		& \citet[]{lafreniere2007}	&	Planet Host	\\
\enddata
\label{table:starproperties}
\end{deluxetable}
\clearpage

\begin{deluxetable}{lccccr}
\tabletypesize{\scriptsize}
\tablewidth{0pt}
\tablecolumns{6}
\tablecaption{Observations Used for RV Pipeline Testing\tablenotemark{a}}
\tablehead{
\colhead{Night} & \colhead{Min. SNR\tablenotemark{b}} &  \colhead{Max. SNR\tablenotemark{b}} & \colhead{Med. SNR\tablenotemark{b}} & \colhead{Total SNR\tablenotemark{b}} & \colhead{$N_{obs}$}
}
\startdata
\multicolumn{6}{c}{GJ 15 A} \\
\hline
2010 Sep 16	&	38	&	49	&	45	&	77	& 	3 \\
2010 Oct 09	&	73	&	78	&	74	&	150	&	4 \\
2010 Oct 10	&	67	&	71	&	69	&	138	&	4 \\
2010 Oct 11	&	32	&	39	&	36	&	89	&	6 \\
2010 Oct 12	&	59	&	64 	&	63	&	152  	&  	6 \\
2010 Oct 13	&	52	&	61 	&	55	&	166	&	9 \\
2010 Nov 22	&	65	&	75 	&	68	&	171	&	6 \\
2010 Nov 23	&	64	&	66 	&	65	&	130	&	4 \\
2010 Nov 24	&	59	&	61 	&	60	&	121	&	4 \\
2011 Jul 10	&	49	&	51 	&	50	&	122	&	6 \\
2011 Jul 13	&	16	&	31 	&	19	&	125	&	30 \\
2011 Aug 19	&	35	&	41 	&	37	&	120	&	10 \\
2011 Aug 20	&	43	&	54 	&	51	&	124	&	6 \\
2012 Dec 10	&	40	&	52 	&	47	&	93	&	4 \\
\hline
\multicolumn{6}{c}{SV Peg} \\
\hline
2010 Sep 15				&	133	&	494	&	316	&   2969	&  83	 \\
2010 Sep 16				&	39	&	299	&	235	&  2047	& 78	\\
2010 Oct 09\tablenotemark{c}\tablenotemark{d}	&	46	&	108	&	86	&	1854	&	480	\\ 
2010 Oct 10\tablenotemark{c}\tablenotemark{d}	&	42	&	106	&	77	&	1162	&	220 	\\ 
2010 Oct 11\tablenotemark{d}				&	14	&	77	&	27	&	356	&	67	\\
2010 Oct 13\tablenotemark{c}\tablenotemark{d}	&	33	&	263	&	75	&	1111	&	112	\\ 
2010 Nov 22\tablenotemark{d}				&	114	&	258	&	175	&	1085	&	36	\\
2010 Nov 23\tablenotemark{d}				&	190	&	216	&	202	&	1395	&	47	\\
2010 Nov 24\tablenotemark{d}				&	104	&	153	& 	122	&	909	&	55	\\
2011 Jul 10\tablenotemark{c}	&	74	&	142	&	115	&	1805	&	246 	\\ 
2011 Aug 19				&	136	&	182	&	162	&	522	&	10	\\
2011 Aug 20\tablenotemark{c}	&	20	&	192	&	56	&	1526	&	602 	\\ 
\hline
\multicolumn{6}{c}{GJ 876} \\
\hline
2010 Sep 15	&	42	&	70	&	58	&	100	&	3 \\
2010 Sep 16	&	45	&	54	&	52	&	101	& 	4 \\
2010 Oct 09	&	42	&	58 	&	56	&	91 	&  	3 \\
2010 Oct 10	&	37	&	42	&	39	&	97	&	6 \\
2010 Oct 11	&	37	&	44	&	42	&	101	&	6 \\
2010 Oct 13	&	32	&	37 	&	34	&	108	&	10 \\
2010 Nov 22	&	36	&	53 	&	52	&	142	&	8 \\
2010 Nov 23	&	38	&	40 	&	40	&	104	&	7 \\
2010 Nov 24	&	40	&	43 	&	41	&	116	&	8 \\
2011 Jul 09	&	32	&	36 	&	33	&	94	&	8 \\
2011 Jul 10	&	30	&	36 	&	36	&	97	&	8 \\
2011 Jul 12	&	27	&	36 	&	33	&	138	&	18 \\
2011 Jul 16	&	31	&	39 	&	32	&	114	&	12 \\
2011 Aug 18	&	54	&	64 	&	58	&	118	&	4 \\
2011 Aug 19	&	60	&	65 	&	61	&	108	&	3 \\
2011 Aug 20	&	42	&	60 	&	51	&	126	&	6 \\
2012 Aug 22	&	50	&	81 	&	80	&	124	&	3 \\
\enddata
\tablenotetext{a}{Taken from the survey conducted by \citet[]{gagne2015}}
\tablenotetext{b}{All SNR values are per pixel}
\tablenotetext{c}{Nights used for intra-night RV stability test}
\tablenotetext{d}{Nights used for intra-run RV stability test}
\label{table:observations}
\end{deluxetable}
\clearpage

\begin{deluxetable}{clll}
\tabletypesize{\scriptsize}
\tablewidth{0pt}
\tablecolumns{4}
\tablecaption{Parameters of the Nominal RV Pipeline Spectral Model}
\tablehead{
\colhead{Parameter} & \colhead{Description} & \colhead{Symbol} & \colhead{Bounded/Variable/Fixed}}
\startdata
1	&	Standard Deviation of Gaussian Term					&	$\omega$			&	Bounded	(0.625 to 1.25 Pixels)\\
2	&	Amplitude of Degree 1 Hermite Polynomial\tablenotemark{a}		&	$a_1$			&	Bounded	($-$0.1 to 0.1)\\
3	&	Amplitude of Degree 2 Hermite Polynomial				&	$a_2$			&	Bounded	($-$0.1 to 0.1)\\
4	&	Amplitude of Degree 3 Hermite Polynomial				&	$a_3$			&	Bounded	($-$0.1 to 0.1)\\
5	&	Amplitude of Degree 4 Hermite Polynomial				&	$a_4$			&	Bounded	($-$0.1 to 0.1)\\
6	&	Coefficient of 0$^{th}$ Order Term in Blaze Function			&	$b_0$			&	Bounded	(0.5 to 2)\\
7	&	Coefficient of 1$^{st}$ Order Term in Blaze Function			&	$b_1$			&	Bounded	($-$0.01 to 0.01)\\
8	&	Coefficient of 2$^{nd}$ Order Term in Blaze Function			&	$b_2$			&	Bounded	($-$0.0001 to 0.0001)\\
9	&	Stellar Doppler Shift\tablenotemark{b} 					&	$\Delta{\ln{\lambda}}$	&	Variable	\\
10	&	Gas Cell Doppler Shift					&	$\Delta \lambda_g$ 	&	Variable	\\
11	&	Telluric Doppler Shift					&	$\Delta \lambda_t$		&	Bounded	($-$2 to 2 \AA)\\
12	&	Gas Cell Optical Depth\tablenotemark{c}				&	$\tau_g$			&	Fixed (Value = 0.97) \\
13	&	Telluric Spectra Optical Depth\tablenotemark{c}				&	$\tau_t$			&	Bounded	(0.4 to 1.5)\\
14	&	Fringing Correction Amplitude\tablenotemark{d}				&	$A$			&	Bounded	(0.001 to 0.05)\\
15	&	Fringing Correction Period\tablenotemark{d} 				&	$B$			&	Bounded	(4 to 8 \AA)\\
16	&	Fringing Correction Phase\tablenotemark{d}				&	$C$			&	Variable \\
17	&	Linear Correction to Wavelength Solution\tablenotemark{e}		&	$\alpha$			&	Bounded	($-$1 to 1 Pixels$^{-1}$)\\
18	&	Quadratic Correction to Wavelength Solution\tablenotemark{e}		&	$\beta$			&	Bounded	($-$1 to 1 Pixels$^{-2}$)\\
\enddata
\tablenotetext{a}{All amplitudes are relative to 1, the amplitude of the 0$^{th}$ degree Hermite Polynomial.}
\tablenotetext{b}{The RVs are calculated by multiplying this parameter by the speed of light.}
\tablenotetext{c}{The optical depth is scaled such that a value of 1 returns the original input spectra.}
\tablenotetext{d}{See Eq. \ref{eq:sinusoid}.}
\tablenotetext{e}{See Eq. \ref{eq:wavelengthscale}.}
\label{table:rvparams}
\end{deluxetable}
\clearpage

\begin{deluxetable}{lrr}
\tabletypesize{\scriptsize}
\tablewidth{0pt}
\tablecolumns{3}
\tablecaption{Nightly Barycenter-Corrected RVs for GJ 15 A , GJ 876, and SV Peg}
\tablehead{
 & \colhead{RV\tablenotemark{a}} &  \colhead{Uncertainty} \\
 \colhead{JD\textendash 2455455\tablenotemark{b}} & \colhead{($\ms$)} & \colhead{($\ms$)}
}
\startdata
\multicolumn{3}{c}{GJ 15 A} \\
\hline
0.834140	&	-35.924	&	 58.905	\\
23.960559	&	-16.589	&	 17.128	\\
24.986080	&	12.105	&	8.653	\\
25.869275	&	-19.026	&	32.705	\\
26.967954	&	-24.966	&	 10.128	\\
27.866246	&	-15.098	&	 19.197	\\
67.907675	&	4.806	&	 14.626	\\
68.825930	&	-25.036	&	 10.005	\\
69.838525	&	15.173	&	 17.589	\\
298.040009	&	-47.045	&	44.270	 \\
301.147817	&	90.232	&	25.418	 \\
339.045267	&	21.583	&	27.680	 \\
339.976246	&	-0.465	&	16.231	 \\
818.876855	&	40.248	&	17.535	 \\
\hline
\multicolumn{3}{c}{SV Peg} \\
\hline
-0.043630	&	-1020.564	&	5.260	 \\
0.959465	&	-971.478 &	4.066	 \\
23.801641	&	-271.968	&	3.071 	 \\
24.902778	&	-238.170	&	4.333	 \\
25.851236	&	-154.422	&	11.137	 \\
27.829183	&	-139.810	&	 5.050	 \\
67.834274	&	1716.403	&	7.937 	 \\
68.814017	&	 1718.879	&	7.166 	 \\
69.829224	&	1808.495	&	7.207 	\\
298.046772	&	-578.908	&	4.842 	 \\
338.941645	&	-900.155	&	16.988 	 \\
339.879140	&	 -968.302	&	3.300 	 \\
\hline
\multicolumn{3}{c}{GJ 876} \\
\hline
-0.017098	&	-274.078	&	9.861	 \\
0.995238	&	-247.526	&	20.746	 \\
23.861299	&	180.486	&	45.491 	 \\
24.927684	&	137.066	&	16.330	 \\
25.821908	&	162.014	&	34.660	 \\
27.815938	&	154.259	&	29.013 	 \\
67.728300	&	29.125	&	11.420 	 \\
68.798500	&	79.733	&	13.565 	 \\
69.778698	&	130.209	&	13.523 	\\
297.134482	&	-244.962	&	21.565 	 \\
298.073901	&	-577.572	&	58.904 	 \\
300.122728	&	-382.672	&	24.414 	 \\
304.006015	&	-367.829	&	22.797 	 \\
337.950553	&	130.333	&	13.222 	 \\
338.956779	&	149.520	&	2.586 	 \\
339.935483	&	157.981	&	28.302 	 \\
706.966665	&	172.149	&	19.817 	 \\
\enddata
\tablenotetext{a}{With mean RV value subtracted}
\tablenotetext{b}{Median of JDs of individual spectra in the given night}
\label{table:gj15a_gj876_svpeg_rv}
\end{deluxetable}
\clearpage

\begin{deluxetable}{lrr}
\tabletypesize{\scriptsize}
\tablewidth{0pt}
\tablecolumns{3}
\tablecaption{Intra-Run Nightly Barycenter-Corrected RVs for SV Peg with Linear Trend Subtracted}
\tablehead{
 & \colhead{RV} &  \colhead{Uncertainty} \\
 \colhead{JD\textendash 2455455\tablenotemark{a}} & \colhead{($\ms$)} & \colhead{($\ms$)}
}
\startdata
\multicolumn{3}{c}{Oct 2010} \\
\hline
23.801641	&	1.518	&	 1.155	\\
24.902778	&	-4.729	&	 1.705	\\
25.851236	&	2.667	&	5.467	\\
27.829183	&	2.362	&	2.432	\\
\hline
\multicolumn{3}{c}{Nov 2010} \\
\hline
67.834274	&	2.988	&	4.064	 \\
68.814017	&	-2.064	&	2.410	 \\
69.829224	&	0.704	&	2.008 	 \\
\enddata
\tablenotetext{a}{Median of JDs of individual spectra in the given night}
\label{table:svpeg_rv}
\end{deluxetable}
\clearpage

\end{document}